\documentclass[aps,prc,amsfonts,preprintnumbers,superscriptaddress,showpacs,nofootinbib]{revtex4}

\usepackage{graphics}
\usepackage{psfrag}
\usepackage{epsfig}
\usepackage{psfig}
\usepackage{epsf}
\usepackage{float}

\newcommand{\fet}[1]{\mbox{\boldmath $#1$}}

\newcommand{\beq}{\begin{equation}}
\newcommand{\eeq}{\end{equation}}
\newcommand{\beqa}{\begin{eqnarray}}
\newcommand{\eeqa}{\end{eqnarray}}
\newcommand{\nn}{\nonumber \\ }

\newcommand{\krig}[1]{\stackrel{\circ}{#1}}

\newcommand{\pr}{\overrightarrow}

\newcommand{\Mp}{M_\pi}

\newcommand{\Mpn}{M_{\pi^0}}

\newcommand{\Mpm}{M_{\pi^\pm}}

\newcommand{\lev}{\overleftarrow}

\begin{document}

\preprint{JLAB-THY-05-302}
\preprint{HISKP-TH-05/05}
\preprint{FZJ-IKP-TH-2005-07}

%\preprint{}

\title{Isospin--violating nucleon--nucleon forces using the method of unitary transformation}


\author{E. Epelbaum}
\email[]{Email: epelbaum@jlab.org}
%\thanks{}
\affiliation{Jefferson Laboratory, Theory Division, Newport News, VA 23606, USA}

\author{Ulf-G. Mei{\ss}ner}
\email[]{Email: meissner@itkp.uni-bonn.de}
\homepage[]{URL: www.itkp.uni-bonn.de/~meissner/}
\affiliation{Universit\"at Bonn, Helmholtz-Institut f{\"u}r
  Strahlen- und Kernphysik (Theorie), D-53115 Bonn, Germany}
\affiliation{Forschungszentrum J\"ulich, Institut f\"ur Kernphysik 
(Theorie), D-52425 J\"ulich, Germany}


\date{\today}

\begin{abstract}
Recently, we have derived the leading and subleading isospin--breaking 
three--nucleon forces using the method of unitary transformation. 
In the present work we extend this analysis and consider the corresponding two--nucleon forces  
using the same approach. Certain contributions to the  isospin--violating one-- and 
two--pion--exchange  potential have already been discussed by various groups within the 
effective field theory framework. Our findings agree with the previously obtained results.
In addition, we present the expressions for the subleading charge--symmetry--breaking two--pion 
exchange potential which were not considered before. These corrections turn out to
be numerically important. 
Together with the three--nucleon force results presented in our previous work, 
the results of the present study 
specify completely isospin--violating nuclear forces up to the order
$q^5/\Lambda^5$, where $q$ ($\Lambda$) denotes the soft (hard) scale. 
\end{abstract}

%\keywords{}

\maketitle

%\section{}
%\subsection{}
%\subsubsection{}

\vspace{-0.2cm}

%%%%%%%%%%%%%%%%%%%%%%%%%%%%%%%%%%%%%%%%%%%%%%%%%%%%%%%%%%%%%%%%%%%%%%%%%%%%%%%%%
\section{Introduction}
\def\theequation{\arabic{section}.\arabic{equation}}
\label{sec:intro}

The interactions between two nucleons are one of the best studied strong
interaction processes, which is due to the large data base of proton-proton
and neutron-proton scattering experiments. As such, many fine features of the
strong interactions and their interplay with the electromagnetic and weak
interactions can be unraveled from such studies, provided a sufficiently
accurate theoretical tool is available to match the sometimes astonishing
precision of the data. In this paper, we are interested in the effects of
isospin violation in the two--nucleon sector. Although isospin is an approximate
symmetry of the QCD Lagrangian, the difference in the up and down quark masses
combined with the long-range electromagnetic force leads to appreciable
deviations from the isospin limit. To quantify these effects, it is mandatory
to have a framework that consistently incorporates these various sources.
Such a scheme has been developed in the last decade, namely chiral nuclear effective
field theory. It extends the so successful chiral perturbation theory for
mesons and meson--baryon systems to processes involving $2,3,4,\ldots$ nucleons. 
Chiral nuclear effective field theory (EFT) has already been applied to study
isospin--violating  two--nucleon forces (2NFs), see e.g. 
\cite{VanKolck:1993ee,vanKolck:1996rm,vanKolck:1997fu,Epelbaum:1999zn,
Walzl:2000cx,Friar:1999zr,Friar:2003yv,Friar:2004ca}. So why coming back to
this topic?  Recently, we have derived the leading and subleading isospin--breaking 
three--nucleon forces (3NFs) using the 
method of unitary transformation \cite{Epelbaum:2004xf}. In the present work, we will 
apply this framework to study the corresponding isospin--breaking two--nucleon forces at the  
same order in the low--momentum expansion. Thereby, we will re--derive the
many interesting results already 
obtained in the earlier studies, but also work out the contributions, which have were never 
been studied before. To the order we are working, we have thus constructed the
complete set of isospin--violating few--nucleon forces.

The material in this paper is organized as follows. In Section~\ref{sec:gen}
we discuss  the effective Lagrangian underlying our calculation and also the
corresponding power counting. Section~\ref{sec:NNforce} is devoted to the
derivation of the isospin--violating 2NFs, consisting of the
one-- and two--pion--exchange potentials and the corresponding contact
interactions. The main novelty is the next--to--leading order
two--pion--exchange potential, which is derived here in its complete form for
the first time and also turns out to be numerically large, very similar to the
isospin--conserving case. In Section~\ref{sec:2N3N} we demonstrate explicitely
the consistency between the two--nucleon forces derived here and the
corresponding isospin--violating three--nucleon forces obtained in 
Ref.~\cite{Epelbaum:2004xf}. Section~\ref{sec:sum}  contains the summary and outlook.

%%%%%%%%%%%%%%%%%%%%%%%%%%%%%%%%%%%%%%%%%%%%%%%%%%%%%%%%%%%%%%%%%%%%%%%%%%%%%%%%%
\section{Power counting and effective Lagrangian}
\def\theequation{\arabic{section}.\arabic{equation}}
\setcounter{equation}{0}
\label{sec:gen}

Within the Standard Model, isospin violation has its origin in 
the different masses of the up and down quarks and the electromagnetic interactions.
At low energy, isospin--breaking effects in few--nucleon systems can be 
studied in the systematic and model--independent framework of  
chiral effective field theory. This method is based on the most general (approximatively) 
chiral invariant Lagrangian for pions and nucleons which 
includes all possible interactions consistent with the isospin violation in the underlying theory.
Consider first isospin--breaking in the strong interactions. 
The QCD quark mass term can be expressed in the two--flavor case as: 
\begin{equation}
\label{epsdef}
{\cal L}_{\rm mass}^{\rm QCD} = -\frac{1}{2}\bar{q} \, 
(m_{\rm u}+m_{\rm d})(1+\epsilon\tau_{3})\,q~, \quad \quad \quad \mbox{where} \quad \quad 
\epsilon \equiv {m_u-m_d \over m_u+m_d} \sim - {1 \over 3}\,.
\end{equation}
The above numerical estimation is based on the light quark mass values
utilizing a modified  $\overline{\rm MS}$ subtraction scheme
at a renormalization scale of 1~GeV \cite{Leutwyler:1996sa}. 
The isoscalar term in Eq.~(\ref{epsdef}) breaks chiral but 
preserves isospin symmetry. It leads to the nonvanishing pion mass, $M^2 =
(m_u + m_d ) B   \neq 0$, 
where $B$  is a low--energy constant (LEC) that describes the strength of the  
bilinear light quark condensate. 
Further, this term generates a string of chiral--symmetry--breaking
interactions in the effective Lagrangian 
which are proportional to positive powers of $M^2$.
The isovector term ($\propto \tau_3$) in Eq.~(\ref{epsdef}) breaks isospin symmetry and generates 
a series of isospin--breaking effective interactions $\propto (\epsilon  M^2)^n$
with $n \geq 1$. It is, therefore, natural to count strong isospin violation in terms of 
$\epsilon  M^2$.\protect\footnote{Notice that  isospin--breaking effects are
in general much smaller than indicated by the numerical value of $\epsilon$, because the
relevant scale for isospin--conserving contributions is the chiral--symmetry--breaking scale 
$\Lambda_\chi$ rather than $m_u+m_d$.}  

Electromagnetic terms in the effective Lagrangian can be generated using the method of external sources, 
see e.g. \cite{Urech:1995aa,Meissner:1997ii,Muller:1999ww} for more details. 
All such terms are proportional to the nucleon charge matrix 
$Q= e \, (1 + \tau_3 )/2$, where $e$ denotes the electric charge.\footnote{Or equivalently, one can use 
the quark charge matrix $e \, (1/3 + \tau_3 )/2$.}
More precisely, the vertices which contain (do not contain) the photon fields are proportional to $Q^{n}$
($Q^{2n}$), where $n=1,2,\ldots$. Since we are interested here in nucleon--nucleon scattering in the absence of 
external fields, so that no photon can leave a Feynman diagram, it is convenient to introduce the 
small parameter $e^2 \sim 1/10$ for isospin--violating effects caused by the electromagnetic interactions. 

In the present study we adopt the same power counting rules for isospin--breaking contributions as specified 
in \cite{Epelbaum:2004xf}. In particular, we count 
\beq\label{CountRules}
\epsilon \sim e \sim \frac{q}{\Lambda}; \quad \quad
\frac{e^2}{(4 \pi )^2}  \sim \frac{q^4}{\Lambda^4}\,,
\eeq
where $q$  ($\Lambda$) refers to a generic low--momentum scale (the pertinent hard scale). 
The $N$--nucleon force receives contributions of the order $\sim (q/\Lambda )^\nu$,
where 
\beq
\label{powc}
\nu = -4 + 2 n_\gamma + 2 N + 2 L + \sum_i V_i \Delta_i\,.
\eeq
Here, $L$ and $V_i$ refer to the number of loops and vertices of type $i$ 
and $n_\gamma$ is the number of virtual photons.
Further, the vertex dimension  $\Delta_i$ is given by
\begin{equation}
\label{chirdim}
\Delta_i = d_i + \frac{1}{2} n_i - 2\;,
\end{equation}
where  $n_i$ is the number of nucleon field operators and $d_i$ is the $q$--power of the 
vertex, which accounts for the number of derivatives and insertions of pion mass, $\epsilon$ and
$e/(4 \pi )$ according to Eq.~(\ref{CountRules}). 
Finally, we adopt the counting rule $q/m \sim (q/ \Lambda )^2$ for the nucleon mass $m$,
which ensures that all iterations of the leading--order NN potential 
contribute to the scattering amplitude at leading order $(q/\Lambda)^0$ and thus have to be resumed, 
see \cite{Weinberg:1990rz,Weinberg:1991um} for more details. 

Let us now specify the terms in the effective Lagrangian we will need in the
present work. It is given in terms of the nucleon isodoublet $N$ and the
isovector pion field $\fet \pi$. Utilizing the heavy baryon framework,
the relevant isospin--symmetric terms in the effective Lagrangian in the
nucleon rest--frame are 
\cite{Bernard:1995dp,Fettes:2001cr}:
\beqa
\label{lagr}
\mathcal{L}^{(0)} &=& \frac{1}{2} \partial_\mu \fet \pi \cdot \partial^\mu \fet \pi  - \frac{1}{2} M^2 \fet \pi^2  
+ N^\dagger  \bigg[  i \partial_0 +
\frac{g_A}{2 F} \fet \tau \vec \sigma \cdot \vec \nabla \fet \pi 
- \frac{1}{4 F^2} \fet \tau \cdot ( \fet \pi \times \dot{\fet \pi } ) \bigg] N  + \ldots  \,, \nn
\mathcal{L}^{(1)} &=& N^\dagger  \bigg[  \delta m +  4 c_1 M^2  
- \frac{2 c_1}{F^2} M^2 \fet \pi^2 + \frac{c_2}{F^2} \dot{\fet \pi}^2  
+ \frac{c_3}{F^2} (\partial_\mu \fet \pi \cdot \partial^\mu \fet \pi )  
%&& \mbox{\hskip 1.25 true cm}  
-  \frac{c_4}{2 F^2} \epsilon_{ijk} \, \epsilon_{abc} \, \sigma_i \tau_a (\nabla_j \, \pi_b ) (\nabla_k \, \pi_c )  
 \bigg] N   + \ldots \,, \nn
\mathcal{L}^{(2)} &=& N^\dagger  \bigg[ \frac{\vec \nabla \, ^2}{2 m} + 
\frac{i g_A}{4 m F} \fet \tau \vec \sigma \cdot ( \lev{\nabla} \dot{\fet \pi} - \dot{\fet \pi} 
\pr{\nabla} ) + \frac{2 d_{16} - d_{18}}{F} M^2  \, \fet \tau \vec \sigma \cdot \vec \nabla \fet \pi 
+ 8 i \tilde d_{28} M^2 \partial_0
\bigg] N  + \ldots \,,
\eeqa
where $M$ denotes the pion mass to leading order in quark masses,  $M =  B
 (m_u + m_d )$,  and $F$ can be identified 
with the pion decay constant in the chiral limit and with the 
electromagnetic interactions being switched off.
Further, $m$ denotes the physical value of the average nucleon mass, $m = (m_p + m_n )/2$, which is related to 
the bare mass $\krig{m}$ via $m = \krig{m} + \delta m$. 
Notice that we are using the physical and not the bare nucleon mass in the heavy baryon expansion,
see \cite{Epelbaum:2002gb} for more details. This leads to the unusual $\delta
m$--term in Eq.~(\ref{lagr}). 
In addition, $g_A$ denotes the axial--vector coupling constant and 
$c_{i}$ and $d_i$ ($\tilde d_i$) are further low--energy constants (LECs). 
The relevant isospin--violating part of the Lagrangian reads \cite{Steininger:1999aa,Fettes:2000gb,Gasser:2002am}:
\beqa
\label{lagrI}
\mathcal{L}^{(2)} &=&  - \frac{e^2}{F^2} C (\fet \pi^2 - \pi_3^2 ) 
+ N^\dagger \Big[2 c_5 \epsilon M^2 \tau_3    
-\frac{c_5}{F^2} \epsilon M^2 (\fet \pi \cdot \fet \tau ) \pi_3 \Big] N  + \ldots \,, \nn [3pt]
\mathcal{L}^{(3)} &=& N^\dagger  \bigg[ f_1 e^2 (\pi_3^2 - \fet \pi^2 ) +  \frac{f_2}{2} e^2 F^2 \tau_3 
+ \frac{f_2}{4} \, e^2 ( (\fet \pi \cdot \fet \tau ) \pi_3 - \fet \pi^2 \tau_3 ) 
%&& \mbox{\hskip 0.7 true cm} 
+ \frac{2 d_{17} - d_{18} - 2 d_{19}}{F} \epsilon 
M^2 \vec \sigma \cdot \vec \nabla \pi_3 \bigg] N  + \ldots\,, \nn [3pt]
\mathcal{L}^{(4)} &=& N^\dagger  \bigg[ \frac{2 g_3 + g_4}{4}\,  e^2 F \, \vec \sigma \cdot \vec \nabla \pi_3
+ \frac{g_4}{4} e^2  F \, \vec \sigma \cdot \vec \nabla \, \pi_3 \, \tau_3 
+ i g_{13}\, e^2 F^2 (1 + \tau_3 ) \partial_0
+ \frac{4 e_{28}}{F} \epsilon M^2 \, \vec \sigma \cdot \vec \nabla \, [\fet \tau \times \dot{\fet \pi } ]_3 \nn
&& \mbox{\hskip 0.7 true cm}  + 8 e_{39} \epsilon M^4 \tau_3 
\bigg] N + \ldots   \,, 
\eeqa
where $C$, $f_i$,  $g_i$ and $e_i$ are further LECs. Here, several comments are in order. 
First, we do not include in Eqs.~(\ref{lagrI}) the $e_{38,40}$-- and $g_{14, 15}$--terms which 
do not lead to isospin--breaking vertices with no pion fields. 
Secondly, for the sake of simplicity, we refrain from showing terms with four  
pion fields in the Lagrangians in Eqs.~(\ref{lagr}), (\ref{lagrI}). The explicit form of such 
terms is of no relevance for our work. We will, however, briefly discuss these terms in section \ref{sec:OPE}. 
In addition, we do not 
show  in Eqs.~(\ref{lagr}), (\ref{lagrI}) isospin--violating NN contact interactions, which 
will be discussed in detail in section \ref{sec:NNcont}. Notice further that the complete form of 
the Lagrangian  $\mathcal{L}^{(5)}$ has not yet been worked out. We will discuss the relevant 
structures from $\mathcal{L}^{(5)}$
in section \ref{sec:OPE}. Finally, we do not consider
terms with photon fields. Those terms give rise to long--range electromagnetic interactions  
between two nucleons, which are extensively discussed in 
Refs.~\cite{Ueling:1935aa,Durand:1957aa,Austin:1983aa,Stoks:1990aa}
and will not be considered in the present work.   
Notice that the effects of these interactions are enhanced at low energy due to their long range, see e.g. \cite{Epelbaum:2004fk}
for more details. In addition to these purely electromagnetic forces, $\pi \gamma$--exchange contributions have to be taken 
into account at the order $\nu = 4$. The explicit expressions for the corresponding NN potential can be found in \cite{vanKolck:1997fu}.

%%%%%%%%%%%%%%%%%%%%%%%%%%%%%%%%%%%%%%%%%%%%%%%%%%%%%%%%%%%%%%%%%%%%%%%%%%%%%%%%%
\section{Isospin--breaking NN force}
\def\theequation{\arabic{section}.\arabic{equation}}
\setcounter{equation}{0}
\label{sec:NNforce}

There are many ways to derive nuclear forces from the effective Lagrangian in Eqs.~(\ref{lagr}), (\ref{lagrI}). 
In this work, we will use the method of unitary transformation  \cite{Epelbaum:1998ka} which leads 
to energy--independent and hermitean potentials. Let us first briefly remind the reader 
on the main idea of this approach. A system of an arbitrary number of interacting pions and nucleons
can be completely described by the Schr\"odinger equation
\begin{equation}
\label{5.1}
H | \Psi \rangle = E | \Psi \rangle~,
\end{equation}
where $H$ denotes the Hamilton operator, which specifies the interaction of pions and nucleons
and can be obtained from the Lagrangian using the canonical formalism.   
Notice that due to the creation of pion field quanta via terms in $H$,
the state $\Psi$ might contain components with an arbitrary 
number of pions. Instead of solving the above infinite--dimensional equation 
for few--nucleon system, it is advantageous to project it onto a  subspace of the 
Fock space, that contains only nucleonic states. 
The resulting equation can be solved using 
the standard methods of few--body physics.
%(XXX too long, was is solved ??).
Let  $\eta$ and $\lambda$  be projection operators on the states 
$| \phi \rangle$ and $| \psi \rangle$
which satisfy $\eta^2 = \eta$, $\lambda^2 = \lambda$, $ \eta \lambda 
= \lambda \eta = 0$ and $\lambda + \eta = {\bf 1}$.
Eq.~(\ref{5.1}) can then be written in the form
\begin{equation}
\label{5.2}
\left( \begin{array}{cc} \eta H \eta & \eta H \lambda \\ 
\lambda H \eta & \lambda  H 
\lambda \end{array} \right) \left( \begin{array}{c} | \phi \rangle \\ 
| \psi \rangle \end{array} \right)
= E  \left( \begin{array}{c} | \phi \rangle \\ 
| \psi \rangle \end{array} \right)
\quad .
\end{equation}
We are now looking for
the unitary operator $U$ which has to be chosen in such a way that the transformed Hamilton
operator is block--diagonal:
\beq
\label{Hbd}
\tilde H \equiv U^\dagger H U = \left( \begin{array}{cc} \eta \tilde H \eta & 0 \\
0 & \lambda \tilde H \lambda \end{array} \right)\,.
\eeq
In \cite{Epelbaum:1998ka} we have adopted the following ansatz for the operator $U$ 
\begin{equation}
\label{5.9}
U = \left( \begin{array}{cc} \eta (1 + A^\dagger A )^{- 1/2} & - 
A^\dagger ( 1 + A A^\dagger )^{- 1/2} \\
A ( 1 + A^\dagger A )^{- 1/2} & \lambda (1 + A A^\dagger )^{- 1/2} \end{array} \right)~,
\end{equation}
which goes back to the work by Okubo \cite{Okubo:1954aa}. The operator $A$ in the above equation 
has only mixed non--vanishing matrix elements: $A = \lambda A \eta$.
We stress that the parameterization of the unitary operator $U$ in Eq.~(\ref{5.9})
is not the most general one. The effective Hamilton 
operator acting on the purely nucleonic subspace of the Fock space can then be obtained 
via
\begin{equation}
\label{n14}
H_{\rm eff} \equiv \eta \tilde H \eta = \eta ( 1 + A^\dagger A)^{-1/2} 
\left( H + A^\dagger H + H A + A^\dagger H A
\right) (1 + A^\dagger A)^{-1/2} \eta~,
\end{equation} 
The requirement in Eq.~(\ref{Hbd}) leads to a set of coupled 
equations for the operator $A$, which can be solved perturbatively  
within the low--momentum expansion along the lines of~\cite{Epelbaum:1998ka}. One then ends up with  
a set of operators that contribute to $H_{\rm eff}$ at a given order in the low--momentum 
expansion. These operators are constructed out of vertices in the effective Lagrangian 
and corresponding energy denominators, see  \cite{Epelbaum:1998ka,Epelbaum:2002gb} for more details. 
The expressions for the nuclear potential are obtained by evaluating two--nucleon (2N), three--nucleon (3N) etc.~matrix elements 
of these operators. Let us now be more specific and consider various isospin--violating 
contributions to the nuclear force up to order $\nu = 5$.

\subsection{One--pion--exchange potential}
\label{sec:OPE}

The isospin--conserving one--pion--exchange (1PE) potential has been studied to one
loop in~\cite{Epelbaum:2002gb} 
using both the S--matrix approach, which relies on the standard technique 
used in quantum field theoretical calculations, 
and the method of unitary transformation. In that work, we restricted
ourselves to isospin--invariant contributions.
We now extend this analysis and include isospin--violating corrections to the 1PE potential up to $\nu = 5$.
While the main focus of \cite{Epelbaum:2002gb} was to study the quark mass dependence of the nuclear force, here we are 
only interested in the physically relevant case and do not need to consider
the chiral expansion of the various LECs. 
Thus, there is no need to evaluate explicitly all loop diagrams
which lead to pion and nucleon mass and wave--function renormalization as well as to renormalization of the pion--nucleon 
vertices. In the following, we will explain how to perform the 
complete calculation of the 1PE potential including renormalization 
of various LECs within the method of unitary transformation and work out the general structure of the 
isospin--breaking 1PE potential up to the considered order.

\begin{figure*}
\vspace{0.5cm}
\centerline{
\psfrag{x11}{\raisebox{-0.2cm}{\hskip 0.0 true cm  (a)}}
\psfrag{x22}{\raisebox{-0.2cm}{\hskip 0.0 true cm  (b)}}
\psfrag{x33}{\raisebox{-0.2cm}{\hskip 0.0 true cm  (c)}}
\psfrag{x44}{\raisebox{-0.2cm}{\hskip 0.0 true cm  (d)}}
\psfrag{x55}{\raisebox{-0.2cm}{\hskip 0.0 true cm  (e)}}
\psfrag{x77}{\raisebox{-0.2cm}{\hskip 0.0 true cm  (f)}}
\psfig{file=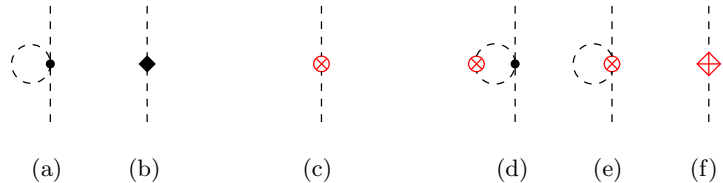,width=10cm}
}
\vspace{0.3cm}
\centerline{\parbox{15.3cm}{
\caption[fig1]{\label{fig1}  Various contributions to the pion mass and wave function renormalization.
Graphs (a) and (b) are the leading isospin--invariant contributions, while the 
diagrams (c) and (d--f) 
show the leading and subleading contributions involving insertions of isospin--violating vertices.  
Dashed lines refer to pions; solid dots and filled diamonds represent isospin--invariant vertices with $\Delta_i =$0 and 2
while crossed circles and crossed diamonds denote isospin--breaking vertices
of dimension  $\Delta_i = $2 and 4, respectively.
}}}
\vspace{0.3cm}
\end{figure*}

First, we introduce, similar to \cite{Epelbaum:2002gb}, renormalized pion fields and masses in the following way:
\beqa
\label{pion}
&& \pi^\pm_{\rm r} = Z_{\pi^\pm}^{-1/2} \, \pi^\pm\,, \mbox{\hskip 1.5 true cm}
Z_{\pi^\pm} = 1 + \delta Z_{\pi}  \,, \mbox{\hskip 2.2 true cm} 
M_{\pi^\pm}^2 = M^2 + \delta  M_\pi^2\,, \nn 
&& \pi^0_{\rm r} = Z_{\pi^0}^{-1/2} \, \pi^0\,, \mbox{\hskip 1.75 true cm}
Z_{\pi^0} = 1 + \delta Z_{\pi} + \delta \bar Z_{\pi} \,, \mbox{\hskip 1.15 true cm} 
M_{\pi^0}^2 = M^2 + \delta M_\pi^2  - \delta \bar  M_\pi^2  \,. 
\eeqa 
The quantities $\delta Z_\pi$ and $\delta  M_\pi^2$ denote isospin--invariant contributions to the 
pion wave--function and $M_\pi^2$ while $\delta \bar  Z_\pi$ and $\delta \bar M_\pi^2$ represent the corresponding isospin--breaking 
terms. These quantities can be expanded in powers of the generic low--momentum
parameters as follows:
\beqa
\label{ChirExp}
&& \mbox{\hskip 0.1 true cm} \delta Z_\pi = \delta  Z_\pi^{(2)} + \delta  Z_\pi^{(4)} + \ldots \,, 
 \mbox{\hskip 2.6 true cm}
\delta \bar Z_\pi = \delta \bar Z_\pi^{(4)} + \ldots \,, \nn
&&  \delta M_\pi^2  = ( \delta M_\pi^2 ) ^{(4)} +
( \delta M_\pi^2 ) ^{(6)} + \ldots  \,, 
 \mbox{\hskip 1.3 true cm}
\delta \bar M_\pi^2  = ( \delta \bar M_\pi^2 ) ^{(4)} + ( \delta \bar M_\pi^2 ) ^{(6)} + \ldots \,,
\eeqa
where superscripts correspond to the power of the small parameters 
according to Eq.~(\ref{CountRules}). 
The leading isospin--invariant corrections $\delta \bar Z_\pi^{(2)}$ and $( \delta \bar M_\pi^2 ) ^{(4)}$ result 
from graphs (a) and (b) in Fig.~\ref{fig1} and have already been considered within the method of unitary transformation in 
\cite{Epelbaum:2002gb}.
Leading and subleading isospin--violating contributions are given by diagrams (c) and (d)--(f) in Fig.~\ref{fig1}, respectively. 
The leading contribution to the charged--to--neutral pion mass difference is entirely of electromagnetic origin and
given by the $C$--term in Eq.~(\ref{lagrI}):
\beq
( \delta \bar M_\pi^2 )^{(4)} =  \frac{2}{F^2} e^2 C\,.
\eeq
Here and in what follows, $F_\pi = 92.4$ MeV refers to the measured value of the (charged) pion decay constant.\footnote{The difference 
between the charged and neutral pion decay constants is $(F_{\pi^\pm} - F_{\pi^0})/F \sim (q/\Lambda )^4$, see 
e.g.~\cite{Steininger:1999aa}. 
Isospin--breaking effects  
due to $ F_{\pi^\pm} \neq F_{\pi^0}$ in the 1PE potential can be accounted for by small shifts in the pion--nucleon coupling constants
as explained below. The corresponding corrections to the 2PE potential enters at order $\nu = 6$ and will not be considered in the 
present work.} 
Notice that graph (c) only contribute to the charged--to--neutral pion mass shift and that there are further pion self--energy corrections due to 
virtual photons, which are not shown explicitly in Fig.~\ref{fig1}. 
The experimentally known pion mass difference $M_{\pi^\pm} - M_{\pi^0} = 4.6$ MeV allows 
to fix the value of the LEC $C$, $C = 5.9 \cdot 10^{-5}$ GeV$^4$. Notice that the natural scale 
for this LEC is $F_\pi^2 \Lambda^2 /(4 \pi)^2 \sim 3 \cdot 10^{-5}$ GeV$^4$ if one adopts $\Lambda \sim M_\rho$.
We do not need to explicitly evaluate higher--order corrections to the pion mass and wave function renormalization 
given by diagrams (d)--(f) in Fig.~\ref{fig1}. The relevant
contributions are incorporated by using physical values for the charged and neutral pion masses, and the  
single--pion Hamilton operator expressed in terms of pion creation and destruction operators $a_i^\dagger$ and $a_i$ 
has the usual form:
\beq
\label{freeHpions}
H_0^\pi = \sum_i \int \frac{d^3 k}{( 2 \pi )^3} \; a_i^\dagger ( \vec k \, ) \, a_i ( \vec k \, ) \; \sqrt{ \vec k \,^2 + M_{i}^2 }\,,
\eeq
where $M_{{1,2}} = M_{\pi^\pm}$, $M_{3} = M_{\pi^0}$.
Notice that at the order considered the effects of the isospin--violating pion wave--function renormalization only shows up 
via an additional interaction 
\beq
\mathcal{H}^{(4)} =  - \frac{g_A \, \delta Z_\pi}{4 F_\pi} \, N^\dagger \tau_3 \,\vec \sigma \cdot \vec \nabla \,  \pi_3 \, N\,,
\eeq
which arises from the $g_A$--vertex being expressed in terms of renormalized pion fields and has the same structure 
as the $g_4$--term in  Eq.~(\ref{lagrI}).
Here and in what follows, we will always work with renormalized pion fields and therefore omit the superscript $\rm r$.

Similar to the pion fields, one can define renormalized proton and neutron fields via $N_p^r = Z_p^{-1/2} \, N_p$,
$N_n^r = Z_n^{-1/2} \, N_n$. In the isospin symmetric case, the leading contribution to $Z_N$ 
results from pion loop and the $\tilde d_{28}$--term in Eq.~(\ref{lagr}) (for 
a detailed discussion of wave function renormalization in the heavy baryon
approach, see Refs.~\cite{Ecker:1997dn},\cite{Steininger:1998ya}).
Clearly, the NN potential derived using the method of unitary 
transformation includes contributions from renormalization of external nucleon lines, which, therefore, do not need to be considered 
separately. Further, we remind the reader that the 
isospin--invariant nucleon mass shift $\delta m$ receives contributions 
at various orders in the low--momentum expansion:
\beq
\delta m = \delta m ^{(1)} + \delta m^{(2)} + \ldots \,, 
\quad \quad \quad \delta m^{(i)} \sim \mathcal{O} \left( \frac{q^{i+1}}{\Lambda^{i}} \right)\,.
\eeq
The leading contribution, $\delta m ^{(1)} = -4 c_1 M_\pi^2$, is clearly due to the second term in the second line of Eq.~(\ref{lagr}) 
while the subleading one, $\delta m ^{(2)}$, receives contributions from pion loops as well as from the counterterms proportional
to LECs $f_{1,2,3}$. In addition to isospin--invariant shifts, there are also isospin--breaking shifts $\delta \bar m$ to the 
nucleon mass $m_N$:
\beq
m_N \equiv \left( \begin{array}{cc} m_p & 0 \\ 0 & m_n \end{array} \right)  = m + \frac{1}{2}\delta \bar m \, \tau_3\,.
\eeq
The leading and subleading contributions to the proton--to--neutron mass difference are of the strong and 
electromagnetic origin, respectively:
\beq
\label{m_shift1}
\delta \bar m ^{(2)} = -4 c_5 \epsilon M_\pi^2 \,, \quad \quad \quad 
\delta \bar m ^{(3)}  = - f_2 \, e^2 F_\pi^2 \,.
\eeq
At the order we are working,
the LECs $c_5$ and $f_2$ can be fixed from the strong and electromagnetic shifts to the nucleon mass:
\beqa
\label{m_shift2}
( m_p - m_n )^{\rm str} &=& (\delta \bar m )^{\rm str} = -2.05 \pm 0.3 \mbox{ MeV ,}  \nn
( m_p - m_n  )^{\rm em} &=& (\delta \bar m )^{\rm em} = 0.76 \pm 0.3  \mbox{ MeV ,} 
\eeqa
which leads to \cite{Meissner:1997ii}
\beq
c_5 = -0.09 \pm 0.01  \mbox{ GeV}^{-1}\,, \quad \quad \quad 
f_2 = -0.45 \pm 0.19    \mbox{ GeV}^{-1}\,.
\eeq
The values for the strong and electromagnetic nucleon mass shifts 
are taken from \cite{Gasser:1982ap}. The electromagnetic shift is based on an evaluation of the
Cottingham sum rule. In principle, this contribution could also be evaluated in chiral perturbation
theory including virtual photons. While the formalism exists (see e.g \cite{Urech:1995aa,Meissner:1997ii,Muller:1999ww}), there
are still some subtleties to be addressed \cite{Gasser:2003hk}. Therefore, we consider the electromagnetic mass
shifts for the ground state baryon octet collected in \cite{Gasser:1982ap} the best values available. 
We stress that there are further corrections to $\delta \bar m$ at higher orders due to pion loop diagrams and counter term 
insertions. We, however, do not need to consider such higher--order corrections in the present work since 
we are interested in isospin--violating corrections to the NN potential up to order $\nu = 5$. Indeed, 
since the isospin--invariant 2PE potential starts to contribute at $\nu = 2$, the corrections 
resulting from insertions of the $\delta \bar m ^{(2)}$-- and $\delta \bar m ^{(3)}$--vertices start to 
contribute at orders $\nu = 4$ and $\nu = 5$, respectively. This can easily be verified using Eq.~(\ref{powc}).
As will be shown below, the corrections to the 1PE potential due to the proton--to--neutron mass difference  are either
proportional to $\delta \bar m /m$ or to $(\delta \bar m )^2$. Consequently, $\delta \bar m ^{(4)}$ first contributes 
to the 1PE potential at order $\nu = 6$ which is beyond the scope of the present work. To the order we are working,
the single--nucleon Hamilton operator takes the form:
\beq
\label{freeH}
H_0^N = \sum_{s, t = \pm 1/2} \int \frac{d^3 p}{(2 \pi )^3} \; n^\dagger_{s, t}  (\vec p \, )  \,  n_{s, t} (\vec p \, )  \;
\bigg[ \frac{\vec p \, ^2}{2 m} + t  \, \delta \bar m  \bigg] \,,
\eeq
where $n^\dagger_{s, t} (\vec p \, )$  ($n_{s, t} (\vec p \, )$) refers to the 
creation (destruction) of a nucleon with 
the spin and isospin quantum numbers $s$ and $t$ and momentum $\vec p$, respectively.
We stress that further isospin--breaking corrections $\propto \delta \bar m / m$ of  kinematical origin have to be taken into account 
in the single--nucleon Hamilton operator entering the Schr\"odinger equation. For our purposes, however, the expression 
for $H_0^N$ in Eq.~(\ref{freeH}) is perfectly sufficient.  
Notice further that here and in what follows, we will not separate the leading and subleading contributions $\delta \bar m ^{(2)}$ and 
$\delta \bar m ^{(3)}$ to the nucleon mass shift: $\delta \bar m \simeq \delta \bar m ^{(2)} +  \delta \bar m ^{(3)}$.

\begin{figure*}
\vspace{0.5cm}
\centerline{
\psfrag{x11}{\raisebox{-0.2cm}{\hskip 0.1 true cm  (a)}}
\psfrag{x12}{\raisebox{-0.2cm}{\hskip 0.1 true cm  (b)}}
\psfrag{x13}{\raisebox{-0.2cm}{\hskip 0.1 true cm  (c)}}
\psfig{file=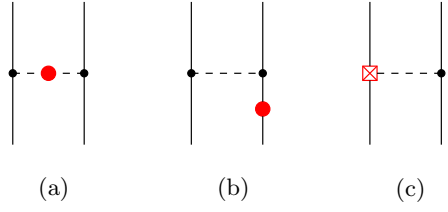,width=6cm}
}
\vspace{0.3cm}
\centerline{\parbox{15cm}{
\caption[fig2]{\label{fig2} Leading ($\nu =2$) and subleading ($\nu =3$) isospin--violating contributions 
to the 1PE potential. Solid lines refer to nucleons and the crossed rectangle 
denotes an isospin--violating 
vertex of dimension $\Delta_i = 3$. A light--shaded circle inserted at a pion
or a nucleon line refers to a single 
insertion of $\delta \bar M_\pi^2$  or $\delta \bar m^{(2)} + \delta \bar m
^{(3)}$, correspondingly. Diagrams
which result from the interchange of the nucleon lines and 
from the application of the time reversal operation
are not shown. For remaining notation see Fig.~\ref{fig1}.
}}}
\vspace{0.3cm}
\end{figure*}

Before going into discussion of various isospin--violating contributions to the  1PE potential, it is instructive to 
recall its general (i.e. without assuming the isospin limit) form based on the 
phenomenological pseudovector (PV)
Lagrangian $\mathcal{L}_{\rm PV}$
which, for instance, in the case of the neutral pion coupled to protons has the form 
\beq
\label{Lpv}
\mathcal{L}_{\rm PV} = \frac{\sqrt{4 \pi} f_{pp\pi^0}}{M_{\pi^\pm}} \,  (\bar N_p i \gamma_\mu \gamma_5 N_p ) \, \partial^\mu \pi^0 \,.
\eeq
Here 
%$N_p$ represents the proton field and 
$f_{pp\pi^0}$ is the corresponding pseudovector coupling constant. 
Similarly, one can define the coupling constants  $f_{nn \pi^0}$, $f_{pn \pi^+}$ and $f_{np \pi^-}$ 
which correspond to the neutral pion coupled to neutrons and charged pions
coupled to nucleons. In
the case of exact isospin symmetry these are related to each other via:
\beq
f_{pp\pi^0} = -  f_{nn\pi^0} = - \frac{1}{\sqrt{2}} f_{pn\pi^+} = \frac{1}{\sqrt{2}} f_{np\pi^-}
\eeq
Notice that 
the charged pion mass enters Eq.~(\ref{Lpv}) just as a scaling factor in order to 
make $f$ dimensionless.  The 1PE potential can then be expressed in a general form as:
\beqa
\label{OPEnijm}
V_{1 \pi} (pp) &=& f_{p}^2 \, V ( M_{\pi^0} )\,, \nn
V_{1 \pi} (nn) &=& f_{n}^2 \, V ( M_{\pi^0} )\,, \nn
V_{1 \pi} (np) &=& - f_{0}^2 \, V ( M_{\pi^0} ) + (-1) ^{I+1}\, 2 f_{c}^2  \, V ( M_{\pi^\pm} )\,, 
\eeqa
where we have introduced the constants $f_p^2 = f_{pp \pi^0} f_{pp \pi^0}$, $f_n^2 = f_{nn \pi^0} f_{nn \pi^0}$, 
$f_0^2 = - f_{pp \pi^0} f_{nn \pi^0}$ and $2 f_c^2 = - f_{np \pi^-} f_{pn \pi^+}$. 
Further, $I=0, 1$ denotes the total isospin of the two--nucleon system and $V (M_{_i} )$ is 
defined as:
\beq
V (M_{i} ) = - \frac{4 \pi}{M_{\pi^\pm}^2} \, \frac {(\vec \sigma_1 \cdot \vec q \,) 
(\vec \sigma_2 \cdot \vec q \,)}{\vec q\, ^2 + M_i^2}\,, 
\eeq
where we use the static approximation for nucleons and neglect all relativistic corrections. 
Charge symmetry implies the same interaction in the {\it pp} and {\it nn} case, i.e. $f_{pp\pi^0} = -  f_{nn\pi^0}$
or $f_p^2 = f_n^2 = f_0^2$. In case of charge independence, one has $f_p^2 = f_n^2 = f_0^2 = f_c^2 \equiv f^2$.
The coupling constant $f^2$ is related to the nucleon axial vector coupling $g_A$ via
\beq
f^2 = \frac{1}{4 \pi} \left[ \frac{g_A M_{\pi^\pm}}{2 F_\pi} \Big( 1 + \delta  \Big) \right]^2\,,
\eeq  
where $\delta$ denotes an isospin--conserving Goldberger--Treiman
discrepancy.\footnote{The Goldberger--Treiman discrepancy is  
defined  in terms of pseudoscalar coupling constant $g$. Pseudoscalar and pseudovector couplings lead to the same 
expression for the 1PE potential on--energy--shell provided the coupling constants are related via $f = g M_{\pi^\pm}/(2 m)$.} 
In the general case 
when isospin symmetry is not conserved, we can introduce in a close analogy to Ref.~\cite{vanKolck:1996rm}
the quantities $\delta_p$, $\delta_n$ and $\delta_c$ corresponding to $f_p^2$,  $f_n^2$ and  $f_c^2$,
respectively. It is sufficient to know the values of these three constants $\delta_p$, $\delta_n$ and $\delta_c$
in order to determine completely the expressions for the 1PE potential in Eqs.~(\ref{OPEnijm}), since the coupling constant 
$f_0^2$ can be expressed as:
 \beq
f^2_0 = \frac{1}{4 \pi} \left[ \frac{g_A M_{\pi^\pm}}{2 F_\pi} \right]^2 \Big( 1 + \delta_p  \Big) \Big( 1 + \delta_n  \Big)\,.
\eeq 
Notice that the dominant contribution to the Goldberger--Treiman discrepancy is generated by the $d_{18}$--term in Eq.~(\ref{lagr})
and does not beak isospin symmetry:
\beq
\delta_p^{(2)} =  \delta_n^{(2)} = \delta_c^{(2)} = - \frac{2 M_\pi^2}{g_A} d_{18} \,.
\eeq
In what follows, we will use the convenient form of the 1PE potential given in Eqs.~(\ref{OPEnijm}) which already incorporates 
the dominant isospin--breaking effects due to the pion--mass difference and charge dependence of the pion--nucleon coupling constant.

We are now in the position to discuss the isospin--violating 1PE potential.
The leading and subleading contributions at orders $\nu =2$ and $\nu =3$ are shown 
in Fig.~\ref{fig2}.  Since we use $\delta \bar m$ and are not separating $\delta \bar m ^{(2)}$ and $\delta \bar m ^{(3)}$, 
graph~(b) in Fig.~\ref{fig2} contains both the order $\nu =2$ and 
$\nu = 3$ contributions to the 1PE potential. Using the method of unitary transformation introduced above and utilizing the notation of 
Refs.~\cite{Epelbaum:1998hg,Epelbaum:2004xf}, one finds the following result for the diagrams (a) and (b) in Fig.~\ref{fig2}:
\beq
\label{OPE}
V_{\rm 1 \pi} = - \frac{1}{2} \eta ' \bigg[ H^{(0)} \frac{\lambda^1}{(H_{0} - E_{\eta})} H^{(0)} 
+   H^{(0)} \frac{\lambda^1}{(H_{0} - E_{\eta '})} H^{(0)} \bigg] \eta \,,
\eeq
where  $\eta$ and $\eta '$ denote the projectors on the purely nucleonic subspace of the Fock space, while 
$\lambda^i$ refers to the projector on the states with $i$ pions. Further, $H^{(0)}$ is the leading 
isospin--invariant $\pi NN$  vertex proportional to $g_A$, 
$H_{0}$ denotes the single--particle Hamilton operator, $H_0 = H_0^\pi + H_0^N$, 
and $E_\eta$  ($E_{\eta '}$) refers to the energy of the nucleons in the state $\eta$ ($\eta '$). 
The contribution of graph (a) to the 1PE potential can be obtained by evaluating the corresponding matrix element 
of the operator $V_{\rm 1 \pi}$ in Eq.~(\ref{OPE}) in the limits $m \to \infty$ and $\delta \bar m \to 0$. 
Clearly, this contribution is already included in Eqs.~(\ref{OPEnijm}).
Similarly,  
the contribution of the diagram (b) is obtained by evaluating the matrix element in the limits $m \to \infty$ and 
$\delta \bar M_\pi^2 \to 0$ and expanding energy denominators in powers of $\delta \bar m$. The term linear in $\delta \bar m$
leads to vanishing matrix elements, so that there is no contribution from graph (b) to the 1PE potential. 
Finally, the contribution of the last diagram (c) can be obtained by evaluating the matrix elements of the operator 
\beq
\label{OPE_2}
V_{\rm 1 \pi} = - \frac{1}{2} \eta ' \bigg[ H^{(0)} \frac{\lambda^1}{(H_{0} - E_{\eta})} H^{(3)}
+   H^{(0)} \frac{\lambda^1}{(H_{0} - E_{\eta '})} H^{(3)} 
\bigg] \eta  + \mbox{h.~c.}\,,
\eeq
in the limits $m \to \infty$, $\delta \bar m \to 0$ and $\delta \bar M_\pi^2 \to 0$. Here, $H^{(3)}$ denotes an
isospin--violating vertex from the Lagrangian $\mathcal{L}^{(3)}$ in Eq.~(\ref{lagrI}) which is 
proportional to the combination of the LECs $2 d_{17} - d_{18} - 2 d_{19}$.  It  
provides a contribution to the quantities $\delta_p$ and $\delta_n$:
\beq
\label{GTDleading}
\delta_p^{(3)} = -  \delta_n^{(3)} = 2 \frac{2 d_{17} - d_{18} - 2 d_{19}}{g_A} \, \epsilon M_\pi^2\,,
\eeq
which leads to the  charge--symmetry breaking 1PE potential.

\begin{figure*}[tbh]
\vspace{0.5cm}
\centerline{
\psfrag{x11}{\raisebox{-0.2cm}{\hskip 0.0 true cm  (a)}}
\psfrag{x12}{\raisebox{-0.2cm}{\hskip 0.0 true cm  (b)}}
\psfrag{x13}{\raisebox{-0.2cm}{\hskip 0.0 true cm  (c)}}
\psfrag{x14}{\raisebox{-0.2cm}{\hskip 0.0 true cm  (d)}}
\psfrag{x15}{\raisebox{-0.2cm}{\hskip 0.05 true cm  (e)}}
\psfrag{x16}{\raisebox{-0.2cm}{\hskip 0.0 true cm  (f)}}
\psfrag{x17}{\raisebox{-0.2cm}{\hskip -0.05 true cm  (g)}}
\psfrag{x18}{\raisebox{-0.2cm}{\hskip 0.0 true cm  (h)}}
\psfrag{x19}{\raisebox{-0.2cm}{\hskip 0.05 true cm  (i)}}
\psfrag{x20}{\raisebox{-0.2cm}{\hskip 0.05 true cm  (j)}}
\psfrag{x21}{\raisebox{-0.2cm}{\hskip 0.0 true cm  (k)}}
\psfrag{x22}{\raisebox{-0.2cm}{\hskip 0.05 true cm  (l)}}
\psfrag{x23}{\raisebox{-0.2cm}{\hskip -0.05 true cm  (m)}}
\psfrag{x24}{\raisebox{-0.2cm}{\hskip 0.05 true cm  (n)}}
\psfrag{x25}{\raisebox{-0.2cm}{\hskip -0.0 true cm  (o)}}
\psfrag{x26}{\raisebox{-0.2cm}{\hskip -0.05 true cm  (p)}}
\psfrag{x27}{\raisebox{-0.2cm}{\hskip 0.05 true cm  (r)}}
\psfrag{x28}{\raisebox{-0.2cm}{\hskip -0.0 true cm  (s)}}
\psfig{file=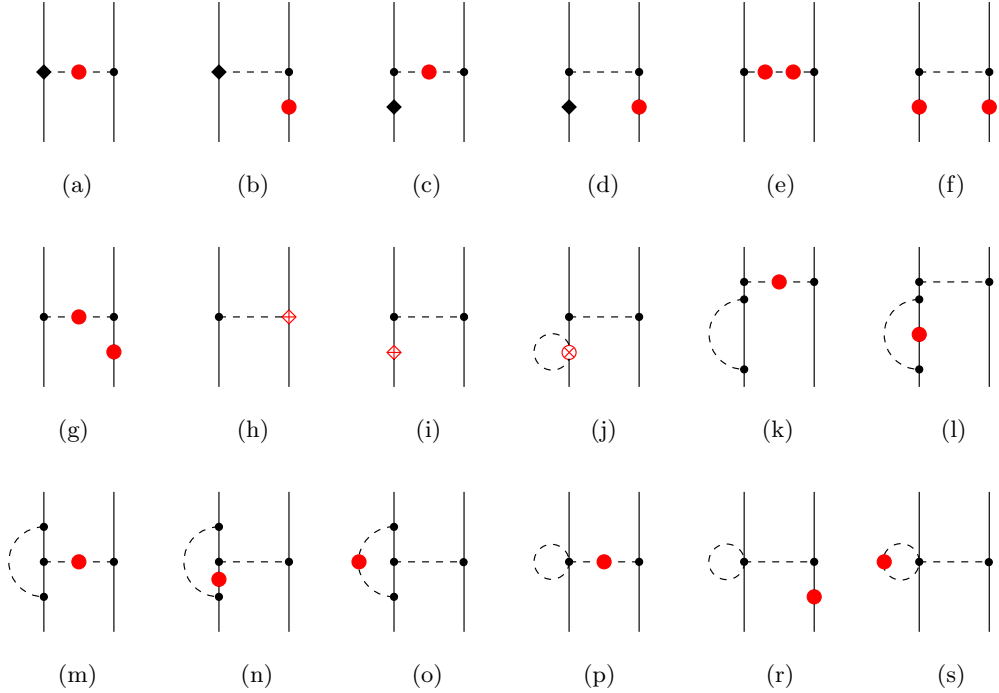,width=14cm}
}
\vspace{0.3cm}
\centerline{\parbox{15cm}{
\caption[fig3]{\label{fig3}   Order $\nu = 4$ contributions to the 1PE potential which 
contain isospin--violating vertices. For diagrams with insertions 
of the nucleon mass shift only one representative topology is depicted. Further graphs 
with nucleon mass insertions at different internal or external nucleon lines are not shown.
For remaining notation, see Figs.~\ref{fig1}, \ref{fig2}.
}}}
\vspace{0.3cm}
\end{figure*}

Isospin--violating contributions to the 1PE potential at order $\nu = 4$ are depicted in Fig.~\ref{fig3}.
Notice that pion tadpole graphs with the Weinberg--Tomozawa vertex lead to vanishing contributions and are not shown   
in Fig.~\ref{fig3}. In addition, pion loop diagrams (but not of the tadpole type) with one insertion of the Weinberg--Tomozawa or $c_5$--vertex 
do not contribute since only odd functions of the loop momentum enter the corresponding integrals. We do not show those
diagrams in Fig.~\ref{fig3} either. Finally, we also ignore one--loop nucleon self--energy contributions with insertions of the pion 
mass difference, since they do not lead to isospin--violating contributions. 
Let us begin with the first graph (a) in Fig.~\ref{fig3}. The contribution proportional to $d_{16}$ 
leads to renormalization of the nucleon axial vector coupling constant $g_A$ while the one proportional to $d_{18}$
provides a dominant contribution to the isospin--conserving Goldberger--Treiman discrepancy, see Eq.~(\ref{GTDleading}). 
Further, the leading relativistic $1/m$--correction vanishes as does the corresponding isospin--invariant contribution.
This can easily be understood in terms of Feynman diagrams. Indeed, the $1/m$--vertex in the last line of Eq.~(\ref{lagr}) 
contains a time derivative of the pion field. Since the four--momentum is conserved, it gives a contribution proportional 
to the nucleon kinetic energy which is of the order $\nu = 6$. For graph (b) we find that the contributions proportional 
to $d_{16}$ and $d_{18}$ vanish. The leading relativistic correction $\propto \delta \bar m / m$ corresponding to this diagram can be 
obtained by evaluating matrix elements of the operators in Eq.~(\ref{OPE_2}), where $H^{(3)}$ corresponds now to the second term 
in the last line of Eq.~(\ref{lagr}). Taking the limit $m \to \infty$ in Eq.~(\ref{freeH}),  expanding the energy denominators in 
powers of $\delta \bar m$ and keeping only terms linear in $\delta \bar m$, we find:
\beq
\label{OPEP2}
V_{\rm 1 \pi}^{3b} = i \frac{\delta \bar m}{2 m} \left( \frac{g_A}{2 F_\pi} \right)^2 
 [ \fet \tau_1 \times \fet \tau_2 ]_3 \, 
\frac{1}{(\vec{q_1} ^2 + M_\pi^2 )} \Big[ (\vec \sigma_1 \cdot \vec q_1) (\vec \sigma_2 \cdot (\vec p_2 + \vec{p_2} ' )) 
+  (\vec \sigma_1 \cdot (\vec p_1 + \vec{p_1} ' ))  (\vec \sigma_2 \cdot \vec q_1) \Big]\,,
\eeq
where $\vec p_i$ ($\vec p_i \, '$) denotes the incoming (outgoing) momentum of the nucleon $i$ and 
$\vec q_1 = \vec{p_1}  ' - \vec p_1 = - (\vec{p_2} ' - \vec{p_2})$.
The order $\Delta_i =2$ isospin--invariant interaction in the next two graphs (c) and (d) is due to the $\tilde d_{28}$--vertex, 
$\delta m^{(2)}$ and the  
nucleon kinetic energy. The $\tilde d_{28}$--term in the Lagrangian is proportional to the nucleon equation of motion and 
is only needed for renormalization purpose.\footnote{It absorbs the ultraviolet divergence in the nucleon Z--factor calculated to one loop.} 
In the method of unitary transformation, this term is eliminated from the Lagrangian by an appropriate field redefinition,
see \cite{Epelbaum:2002gb} for more details. 
The contributions from graphs (c) and (d) proportional to the 
nucleon kinetic energy can be obtained evaluating matrix elements 
of the operators in Eq.~(\ref{OPE}), expanding in $\delta \bar M_\pi^2$, $1/m$ and $\delta \bar m $ and keeping only terms proportional 
to $\delta \bar M_\pi^2/m$ and $\delta \bar m /m$, respectively. 
We find that graph (c) leads to a vanishing result while graph (d) provides 
the following contribution to the 1PE potential:
\beq
\label{OPEP1}
V_{1 \pi}^{3d} = - i \frac{\delta \bar m}{2 m} \left( \frac{g_A}{2 F_\pi} \right)^2  [ \fet \tau_1 \times \fet \tau_2 ]_3 \,
\frac{(\vec \sigma_1 \cdot \vec q_1\, )(\vec \sigma_2 \cdot \vec q_1 \,)}{(\vec{q_1} ^2 + M_\pi^2 )^2} 
(\vec{p_1}^2 - \vec{p_2}^2 - {\vec{p_1} '}{}^2 + {\vec{p_2} '}{}^2 )\,,
\eeq
Introducing the total momentum $\vec P$
of the two--nucleon system, $\vec P = \vec p_1 + \vec p_2 =  \vec{p_1}' + \vec{p_2} '$, 
the above expression can be re--written as:
\beq
V_{1 \pi}^{3d} = i \frac{\delta \bar m}{m} \left( \frac{g_A}{2 F_\pi} \right)^2  [ \fet \tau_1 \times \fet \tau_2 ]_3 \,
\frac{(\vec \sigma_1 \cdot \vec q_1\, )(\vec \sigma_2 \cdot \vec q_1 \,)}{(\vec{q_1} ^2 + M_\pi^2 )^2} 
\, \vec q_1 \cdot \vec P\,.
\eeq
Thus, this potential vanishes in the two--nucleon center--of--mass (CMS). 
The contributions from graphs (c) and (d) proportional to $\delta m^{(2)}$ are found to vanish.
As pointed out before, the contribution from diagram (e) in Fig.~\ref{fig3} proportional to $(\delta \bar M_\pi^2 )^2$ is already 
taken into account in Eqs.~(\ref{OPEnijm}). For graph (f) we obtain the following contribution to the 1PE potential:
\beq
\label{OPEP3}
V_{\rm 1 \pi}^{3f} = - (\delta \bar m)^2 \left( \frac{g_A}{2 F_\pi} \right)^2 
\left( \fet \tau_1 \cdot \fet \tau_2  - \tau_1^3 \tau_2^3 \right) \, 
\frac{(\vec \sigma_1 \cdot \vec q\, )(\vec \sigma_2 \cdot \vec q \,)}{(\vec{q} \,^2 + M_\pi^2 )^2} \,.
\eeq
Notice that the isospin--violating piece has the same structure as the correction due to the pion mass difference at order $\nu = 2$ but is 
$\delta \bar M_\pi^2 / (\delta \bar m)^2 \sim 660$ times weaker. 
Next, graph (g) is expected to provide a correction proportional to $\delta \bar m \, \delta \bar M_\pi^2$. We find that the 
corresponding contribution vanishes. Further, the contribution of diagram (h) is included in the 1PE potential in Eqs.~(\ref{OPEnijm}),
where one has to account for shifts in the pion--nucleon coupling constants:
\beq
\delta f_p^{(4)} =  \frac{g_3 + g_4}{g_A} \, e^2 F_\pi^2 + \frac{\delta Z_\pi}{2} \,, \quad \quad
\delta f_n^{(4)} =  - \frac{g_3}{g_A} \, e^2 F_\pi^2  + \frac{\delta Z_\pi}{2}  \,.
\eeq
Consider now diagram (i) in Fig.~\ref{fig3}. The isospin--violating
counterterms of dimension $\Delta_i =4$ include both the strong 
term $\propto e_{39}$ which leads to the nucleon mass shift, and the 
electromagnetic one $\propto g_{13}$ which is proportional to the nucleon equation of motion and absorbs the ultraviolet 
divergence in the nucleon $Z$--factor. We find that this diagram as well as \emph{all} remaining diagrams (j)--(s) in 
Fig.~\ref{fig3} either lead to vanishing contributions or renormalize various LECs. 
In particular, contributions of graphs (k) and (i), (j), (l) can be expressed in terms of isospin--conserving and isospin--violating 
nucleon self--energy corrections, while diagrams (m), (p), (r) and (n), (o), (s) give rise to isospin--invariant and isospin--breaking 
renormalization of the pion--nucleon coupling constants. None of the pion loop diagrams lead to any form--factor--like behavior, i.e. have 
a non--trivial dependence on the momentum transfer between two nucleons. Thus the contribution of all these diagrams is taken into account 
by using the general expressions for the one--pion exchange potential in Eqs.~(\ref{OPEnijm}) expressed in terms of renormalized quantities. 
Stated differently, the contributions of these diagrams only lead to charge--dependent shifts in the strength of the 1PE potential. 
We stress that since the corresponding LECs $g_3^{\rm r}$ and $g_4^{\rm r}$ are not known experimentally and have to be determined 
from the data, and because we are not interested in the quark--mass dependence of the strength of the 1PE potential, 
we do not need to evaluate the loop diagrams in Fig.~\ref{fig3} explicitly. 
Finally, we have also verified the finding of Ref.~\cite{vanKolck:1996rm}, that the contributions
of diagrams (l) and (n) cancel.

\begin{figure*}[tb]
\vspace{0.5cm}
\centerline{
\psfrag{x11}{\raisebox{-0.2cm}{\hskip 0.0 true cm  (a)}}
\psfrag{x12}{\raisebox{-0.2cm}{\hskip 0.0 true cm  (b)}}
\psfrag{x13}{\raisebox{-0.2cm}{\hskip 0.0 true cm  (c)}}
\psfrag{x14}{\raisebox{-0.2cm}{\hskip 0.0 true cm  (d)}}
\psfrag{x15}{\raisebox{-0.2cm}{\hskip 0.05 true cm  (e)}}
\psfrag{x16}{\raisebox{-0.2cm}{\hskip 0.0 true cm  (f)}}
\psfrag{x17}{\raisebox{-0.2cm}{\hskip -0.05 true cm  (g)}}
\psfrag{x18}{\raisebox{-0.2cm}{\hskip 0.0 true cm  (h)}}
\psfrag{x19}{\raisebox{-0.2cm}{\hskip 0.05 true cm  (i)}}
\psfrag{x20}{\raisebox{-0.2cm}{\hskip 0.05 true cm  (j)}}
\psfrag{x21}{\raisebox{-0.2cm}{\hskip 0.0 true cm  (k)}}
\psfrag{x22}{\raisebox{-0.2cm}{\hskip 0.05 true cm  (l)}}
\psfrag{x23}{\raisebox{-0.2cm}{\hskip -0.05 true cm  (m)}}
\psfrag{x24}{\raisebox{-0.2cm}{\hskip 0.05 true cm  (n)}}
\psfrag{x25}{\raisebox{-0.2cm}{\hskip -0.0 true cm  (o)}}
\psfrag{x26}{\raisebox{-0.2cm}{\hskip 0.0 true cm  (p)}}
\psfrag{x27}{\raisebox{-0.2cm}{\hskip 0.05 true cm  (q)}}
\psfrag{x28}{\raisebox{-0.2cm}{\hskip -0.05 true cm  (r)}}
\psfrag{x29}{\raisebox{-0.2cm}{\hskip 0.05 true cm  (s)}}
\psfrag{x30}{\raisebox{-0.2cm}{\hskip -0.0 true cm  (t)}}
\psfrag{x31}{\raisebox{-0.2cm}{\hskip 0.05 true cm  (u)}}
\psfrag{x32}{\raisebox{-0.2cm}{\hskip -0.0 true cm  (v)}}
\psfig{file=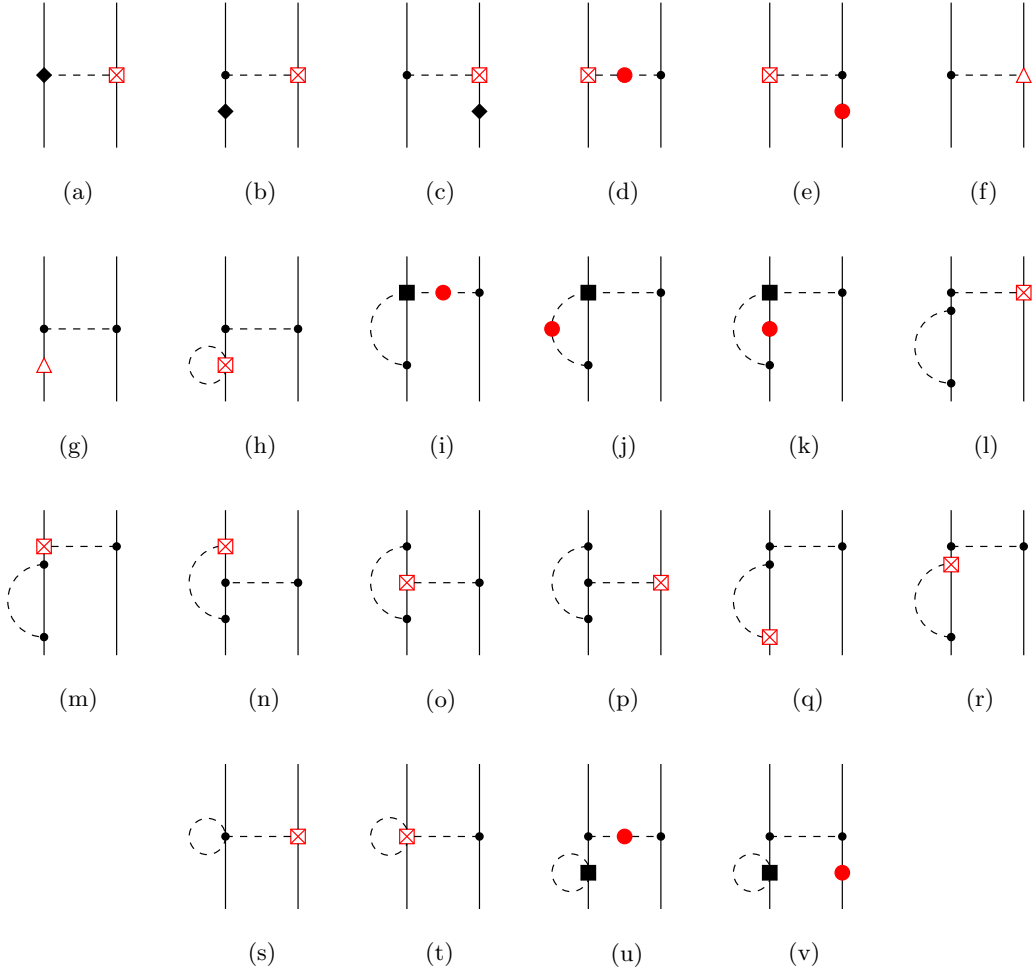,width=14cm}
}
\vspace{0.3cm}
\centerline{\parbox{15cm}{
\caption[fig4]{\label{fig4}   Order $\nu = 5$ contributions to the 1PE potential which 
contain isospin--violating vertices. Filled rectangles refer to isospin--invariant vertices of dimension 
$\Delta_i = 1$ while triangles denote isospin--breaking interactions of dimension $\Delta_i = 5$.
For remaining notation, see Figs.~\ref{fig1}, \ref{fig2}, \ref{fig3}. 
}}}
\vspace{0.3cm}
\end{figure*}

Let us now discuss the corrections to the 1PE potential at order $\nu = 5$
represented by 
the diagrams depicted in Fig.~\ref{fig4}.
Notice that we again refrain from showing various kinds of diagrams which lead to vanishing contributions as explained above. 
For the same reason, we also do not show graphs with an insertion of the $\pi \pi NN$ vertices proportional to $f_{1}$, $f_2$ 
(but not of the tadpole type) as well as 
pion tadpole diagrams proportional to $c_i$ with an insertion of the $\delta \bar M_\pi^2$--vertex at the pion line, which do not 
lead to isospin--violating contributions to the potential.
First, we note that the contribution of graph (a) proportional to the LECs $d_{16}$ and $d_{18}$ is already included in 
Eqs.~(\ref{OPEnijm}). Further, similar to the case of diagram (a) in Fig.~\ref{fig3}, the corresponding $1/m$--correction vanishes 
at this order. Diagrams (b) and (c) also do not lead to new structures in the 1PE potential: the term $\propto \tilde d_{28}$
provides a shift of the strength of the 
charge--symmetry--breaking (CSB) 1PE potential 
via the nucleon wave--function renormalization while the contributions
proportional to the nucleon kinetic energy and to $\delta m^{(2)}$ vanish.
Similarly, diagrams of the type (c) and (d) in Fig.~\ref{fig3} but proportional to $\delta m^{(3)}$ instead of 
$\delta m^{(2)}$, which are not shown explicitly in Fig.~\ref{fig4}, lead to a
vanishing result. 
The contribution of diagram (d) is already included in the expression 
for the 1PE potential in Eqs.~(\ref{OPEnijm}). Graph (e) leads to a vanishing
result.  Consider now the contribution of diagram (f). 
Order $\Delta_i = 5$ isospin--violating counter terms have not yet been worked out. Here we simply list all possible  
structures consistent with the usual symmetry constraints: 
\beq
\mathcal{L}^{(5)} = N^\dagger \Big( a \, e^2 \, \vec \sigma \cdot \vec \nabla \, [ \fet \tau \times \dot{\fet \pi } ]_3 
+ b_1 \, \epsilon M_\pi^2 \, \vec \sigma \cdot \vec \nabla \,  \vec \nabla ^2 \pi_3   + b_2 \, \epsilon M_\pi^2 
\vec \sigma \cdot \vec \nabla \ddot{\pi}_3  + b_3 \, \epsilon M_\pi^4 \, \vec \sigma \cdot \vec \nabla \pi_3 \Big) N\,,  
\eeq
where $a$ and $b_i$ are LECs. The first term in the above expression does not contribute at order $\nu = 5$
due to the presence of the time derivative. The 1PE potential proportional to $b_i$ can be cast into the form 
of Eqs.~(\ref{OPEnijm}) plus additional CSB short--range interactions. 
Further, one has to take into account relativistic $1/m$--terms
\beq
\mathcal{L}^{(5)} = i \frac{2 d_{17} - d_{18} - 2 d_{19}}{2 m F_\pi} \epsilon M_\pi^2 \; N^\dagger 
\vec \sigma \cdot ( \lev{\nabla} \dot{\pi}_3 - \dot{\pi}_3 
\pr{\nabla} )  N \,,
\eeq
which, however, contribute at higher orders due to the presence of the time derivative. 
Graph (g) in Fig.~\ref{fig4} represents the correction to the 1PE potential due to an insertion 
of the order $\Delta_i = 5$ isospin--violating counter terms. To the best of our knowledge, the latter 
have not yet been worked out. They may include $\epsilon M_\pi^4$--terms proportional to the nucleon 
equation of motion which contribute to the nucleon Z--factor as well as terms $\propto M_\pi^2 \, e^2/(4 \pi)^2$
which give further corrections to the nucleon mass difference. Again, we do not need to evaluate explicitly 
the contributions of this diagram as well as all remaining graphs (h)--(v)  
in  Fig.~\ref{fig4} since they 
do not lead to any new structures in the 1PE potential. We have verified that diagrams of type (k)
only lead to shifts in the pion--nucleon coupling constants $\delta_i$. We further note that the loop integral in 
graph (j) has no logarithmic ultraviolet divergence, which is consistent with the fact that diagram (f) has no contribution  
due to counter terms of electromagnetic origin. 

To summarize, isospin--violating corrections to the 1PE potential up to $\nu = 5$  are accounted for 
by using the expression in Eqs.~(\ref{OPEnijm}) and further corrections in Eqs.~(\ref{OPEP2}), (\ref{OPEP1})
and (\ref{OPEP3}). The CSB corrections in Eqs.~(\ref{OPEP2}) and (\ref{OPEP1}) obtained within the method 
of unitary transformation agree with the ones found in \cite{Friar:2004ca} using a completely different framework.
%*EE
The correction $\propto (\delta \bar m)^2$ in Eq.~(\ref{OPEP3}) 
%was not mentioned in that work. It 
can be found e.g.~in \cite{Stoks:1990bb}, see also discussion in \cite{Friar:2004ca}.

\subsection{Two--pion--exchange potential}
\label{sec:TPE2N}

\begin{figure*}[tb]
\vspace{0.5cm}
\centerline{
\psfrag{x11}{\raisebox{-0.0cm}{\hskip 0.1 true cm  (a)}}
\psfrag{x12}{\raisebox{-0.0cm}{\hskip 0.1 true cm  (b)}}
\psfrag{x13}{\raisebox{-0.0cm}{\hskip 0.1 true cm  (c)}}
\psfrag{x14}{\raisebox{-0.0cm}{\hskip 0.1 true cm  (d)}}
\psfrag{x15}{\raisebox{-0.0cm}{\hskip 0.1 true cm  (e)}}
\psfrag{x16}{\raisebox{-0.0cm}{\hskip 0.1 true cm  (f)}}
\psfrag{x17}{\raisebox{-0.0cm}{\hskip 0.1 true cm  (g)}}
\psfrag{x18}{\raisebox{-0.0cm}{\hskip 0.1 true cm  (h)}}
\psfrag{x19}{\raisebox{-0.0cm}{\hskip 0.1 true cm  (i)}}
\psfrag{x20}{\raisebox{-0.0cm}{\hskip 0.1 true cm  (j)}}
\psfig{file=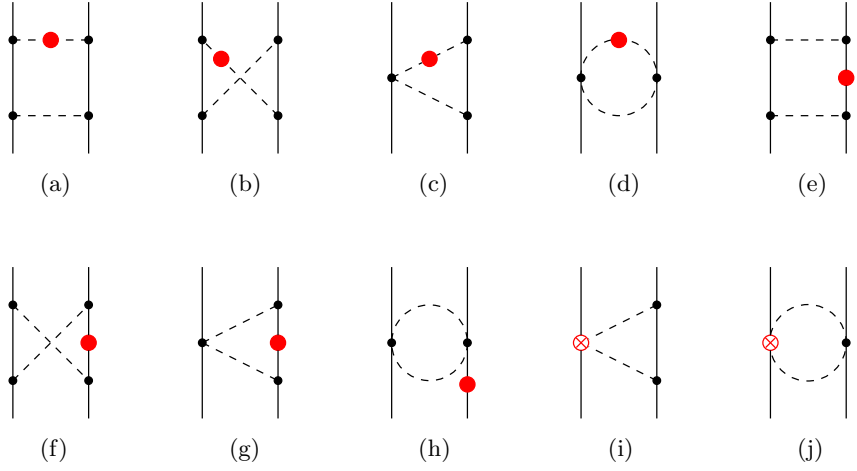,width=12cm}
}
\vspace{0.3cm}
\centerline{\parbox{14cm}{
\caption[fig5]{\label{fig5} Leading isospin--breaking correction to the 2PE potential at order $\nu = 4$.
For notation, see Figs.~\ref{fig1}, \ref{fig2} and \ref{fig3}.
}}}
\vspace{0.3cm}
\end{figure*}

Let us now discuss the leading and subleading isospin--violating two--pion--exchange potential which 
can be expressed in momentum space as:
\beq 
\label{2PEmom_gen}
V_{2 \pi} = (\tau_1^3 \, \tau_2^3 ) \, \Big[ V_C + V_S (\vec \sigma_1 \cdot \vec \sigma_2 ) + 
V_T (\vec \sigma_1 \cdot \vec q \, )  (\vec \sigma_2 \cdot \vec q \, ) \Big] +
(\tau_1^3 +\tau_2^3 ) \, \Big[ W_C + W_S (\vec \sigma_1 \cdot \vec \sigma_2 ) + 
W_T (\vec \sigma_1 \cdot \vec q \, )  (\vec \sigma_2 \cdot \vec q \, ) \Big]\,,
\eeq
with the six functions $V_C (q), \ldots , W_T (q)$ depending on the momentum transfer $q \equiv | \vec q \, |$. 
The subscripts refer to the central (C), spin--spin (S) and tensor (T)
components in the potential. Further, 
$V_i $ and $W_i$ correspond to charge--symmetry conserving and charge--symmetry breaking pieces, respectively. 
The dominant contributions 
arise at order $\nu = 4$ from diagrams shown in Fig.~\ref{fig5}. In this figure, graphs (a)--(d) represent 
the effects due to the pion mass difference, graphs (e)--(h) provide contributions proportional to the 
nucleon mass difference, and the last two graphs (i) and (j) are due to a single insertion of the isospin--breaking 
$\pi \pi NN$--vertex proportional to the LEC $c_5$. 

The isospin--violating two--pion--exchange (2PE) 
potential due to the pion mass difference has been considered in 
\cite{Friar:1999zr}. As shown in this reference, it can be expressed in terms of the corresponding 
isospin--invariant contributions without performing any additional calculations. To that aim, one can first 
decompose the isospin--invariant 2PE potential into the isoscalar and isovector pieces
\beq
V_{2\pi} = V_{2\pi}^0 + V_{2\pi}^1 \, \fet {\tau}_1 \cdot
\fet {\tau}_2~.
\eeq
The leading isospin--breaking effects die to $M_{\pi^\pm} \neq M_{\pi^0}$  are  incorporated properly if 
one uses $\tilde M_\pi$, defined as 
\beq\label{avmass}
\tilde \Mp = \frac{2}{3} \Mpm +  \frac{1}{3} \Mpn~,
\eeq
in the scalar part $V_{2\pi}^0$ and expresses the vector part as following:
\beq
V_{2\pi}^1 = \left\{ \begin{array}{ll}  V_{2\pi}^1  (M_{\pi^\pm} )     & {\rm for}~~pp~~{\rm and}~~nn~, \\[1ex]
                                        V_{2\pi}^1  (M_{\pi^0} )       & {\rm for}~~np,~~T = 1~, \\[1ex]
                                        V_{2\pi}^1  (\tilde M_{\pi} )  & {\rm for}~~np,~~T = 0~. \end{array}\right.
\eeq
These results are valid  modulo $( \delta \bar M_\pi^2 /M_\pi^2 )^2$--corrections.
Notice that terms $\propto ( \delta \bar M_\pi^2 /M_\pi^2 )^2$  start to contribute 
to the 2PE potential at order $\nu = 6$ and thus need not be considered in the present work. 
Equivalently, we can write the 2PE potential in the form
\beq
\label{2PECIB}
V_{2\pi} = V_{2\pi}^0 (\tilde M_\pi ) + V_{2\pi}^1 (\bar M_\pi ) \;  \fet {\tau}_1 \cdot \fet {\tau}_2
+ \frac{\delta \bar M_\pi^2}{4 M_\pi } \; \frac{\partial V_{2\pi}^1 (M_\pi )}{\partial M_\pi } \; \tau_1^3 \, \tau_2^3
+ \mathcal{O} \left( \left( \frac{\delta M_\pi^2}{M_\pi^2} \right)^2 \right)\,,
\eeq
where $\bar M_\pi$ is the average pion mass 
\beq
\bar M_\pi = \frac{1}{2} \left( \Mpm +  \Mpn \right) \,.
\eeq
Substituting the expressions for the isospin--invariant 2PE potential at $\nu = 2$ given e.g.~in \cite{Epelbaum:2003gr}
into Eq.~(\ref{2PECIB}) we obtain for the non--polynomial parts of the contribution of diagrams (a)--(d) in Fig.~\ref{fig5}:
\beq
\label{VC4}
V_{C}^{(4)} = \frac{\delta \bar M_\pi^2}{128 \pi^2 F_\pi^4}  \;
\frac{1}{4 M_\pi^2 + q^2} \, \bigg\{ 4 g_A^4 M_\pi^2 
%\nn && {} 
- \bigg( 4 M_\pi^2 ( 9 g_A^4 - 4 g_A^2 - 1 ) + q^2 (11 g_A^4 - 6 g_A^2 - 1) - \frac{16 g_A^4 M_\pi^4}{4 M_\pi^2 + q^2} \bigg) 
\, L^\Lambda (q) \bigg\}\,,
\eeq
where the loop function $L^\Lambda (q)$ reads 
\beq
\label{def_LA}
L^{\Lambda} (q) =  \frac{\omega}{2 q} \, 
\ln \frac{ \Lambda^2 \omega^2 + q^2 s^2 + 2  \Lambda q 
\omega s}{4 M_\pi^2 (  \Lambda^2 + q^2)}~, \quad \quad
\omega = \sqrt{ q^2 + 4 M_\pi^2}~,  \quad \quad
s = \sqrt{ \Lambda^2 - 4 M_\pi^2}\,.
\eeq
The above expression for $L^\Lambda (q)$ is given in the spectral--function
regularization (SFR) 
framework with
$\Lambda$ being the corresponding cut--off. The limit $\Lambda \to \infty$
corresponds to dimensional regularization (DR).  

\begin{figure*}[tb]
\vspace{0.5cm}
\centerline{
\psfrag{x11}{\raisebox{-0.0cm}{\hskip 0.1 true cm  (a)}}
\psfrag{x12}{\raisebox{-0.0cm}{\hskip 0.1 true cm  (b)}}
\psfrag{x13}{\raisebox{-0.0cm}{\hskip 0.1 true cm  (c)}}
\psfrag{x14}{\raisebox{-0.0cm}{\hskip 0.1 true cm  (d)}}
\psfrag{x15}{\raisebox{-0.0cm}{\hskip 0.1 true cm  (e)}}
\psfrag{x16}{\raisebox{-0.0cm}{\hskip 0.1 true cm  (f)}}
\psfrag{x17}{\raisebox{-0.0cm}{\hskip 0.1 true cm  (g)}}
\psfrag{x18}{\raisebox{-0.0cm}{\hskip 0.1 true cm  (h)}}
\psfrag{x19}{\raisebox{-0.0cm}{\hskip 0.1 true cm  (i)}}
\psfrag{x20}{\raisebox{-0.0cm}{\hskip 0.1 true cm  (j)}}
\psfig{file=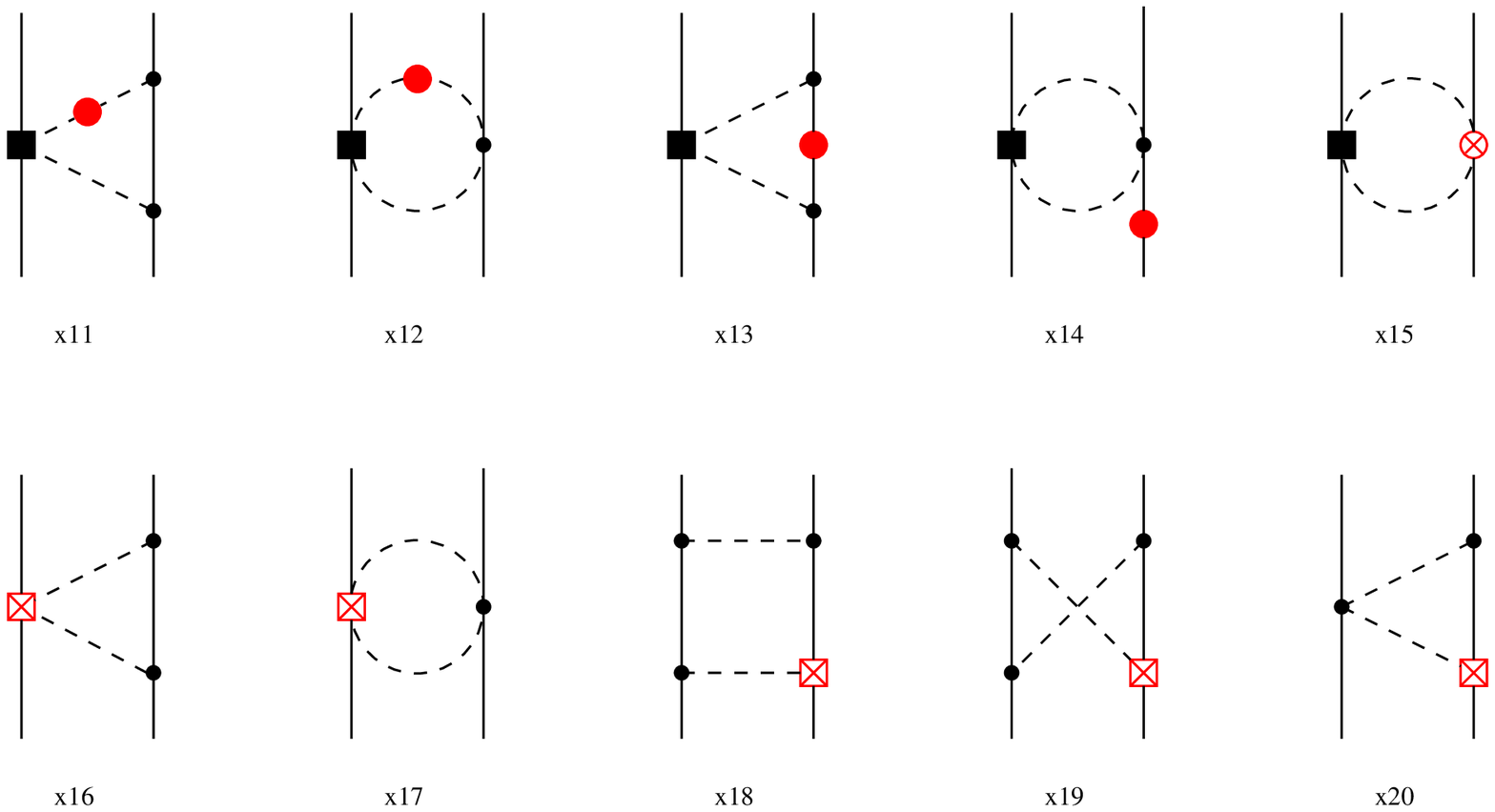,width=12cm}
}
\vspace{0.3cm}
\centerline{\parbox{14cm}{
\caption[fig6]{\label{fig6} Subleading isospin--breaking correction to the 2PE potential 
at order $\nu = 5$.   For notation see Figs.~\ref{fig1}--\ref{fig4}.
}}}
\vspace{0.3cm}
\end{figure*}

Consider now diagrams (e)--(h) in Fig.~\ref{fig5} which include one insertion of the nucleon mass shift. 
The contributions of graphs (e) and (f) can be obtained within the method of unitary transformation 
by evaluating the corresponding matrix elements of the operator
\beqa
\label{5e5f}
V_{2 \pi}^{\rm 5e,  5f} &=& \eta ' \bigg[ \frac{1}{2} H^{(0)} \frac{\lambda^1}{(H_0 - E_{\eta '})}  H^{(0)} \, \tilde \eta 
\,  H^{(0)} \frac{\lambda^1}{(H_0 - E_{\tilde \eta} )( H_0 - E_{\eta '} )}  H^{(0)} \nn 
&& \mbox{\hskip 0.7 true cm} -\frac{1}{8} H^{(0)} \frac{\lambda^1}{(H_0 - E_{\eta '})}  H^{(0)} \, \tilde \eta 
\,  H^{(0)} \frac{\lambda^1}{(H_0 - E_{\tilde \eta} )( H_0 - E_{\eta} )}  H^{(0)} \nn
&&  \mbox{\hskip 0.7 true cm} + \frac{1}{8} H^{(0)} \frac{\lambda^1}{(H_0 - E_{\eta '}) ( H_0 - E_{\tilde \eta} )}  
H^{(0)} \, \tilde \eta 
\,  H^{(0)} \frac{\lambda^1}{(H_0 - E_{\tilde \eta} )}  H^{(0)} \nn
&&  \mbox{\hskip 0.7 true cm} - 
\frac{1}{2} H^{(0)} \frac{\lambda^1}{(H_0 - E_{\eta})}  H^{(0)} \,  \frac{\lambda^2}{(H_0 - E_{\eta})} 
\,  H^{(0)} \frac{\lambda^1}{(H_0 - E_{\eta} )}  H^{(0)}
\bigg] \eta  + \mbox{h.~c.}
\eeqa
in the limits $m \to \infty$, $\delta \bar M_\pi^2 \to 0$ by expanding the denominators in powers of $\delta \bar m$ 
and keeping only terms linear in  $\delta \bar m$.\footnote{Notice that the contribution $\propto \delta \bar M_\pi$ 
discussed above can also be obtained from Eq.~(\ref{5e5f}) if one takes the limits $m \to \infty$, $\delta \bar m \to 0$
in $H_0$. }
Similarly, contributions of diagrams (g) and (h) can be obtained by evaluating the matrix elements of the operators
\beqa
\label{5g}
V_{2 \pi}^{\rm 5g} &=& \frac{1}{2} \eta ' \bigg[ H^{(0)} \frac{\lambda^1}{(H_0 - E_{\eta})}  H^{(0)} 
\frac{\lambda^2}{(H_0 - E_{\eta})}
H^{(0)} + H^{(0)} \frac{\lambda^2}{(H_0 - E_{\eta})} H^{(0)} \frac{\lambda^1}{(H_0 - E_{\eta})}  H^{(0)} \nn
&&{} \mbox{\hskip 0.8 true cm} + H^{(0)} \frac{\lambda^1}{(H_0 - E_{\eta}) }  H^{(0)}   
\frac{\lambda^1}{(H_0 - E_{\eta}) } H^{(0)} \bigg] \eta  + \mbox{h.~c.}\,,
\eeqa
and 
\beq
V_{2 \pi}^{\rm 5h} = \frac{1}{2} \eta '  \bigg[  H^{(0)} \frac{\lambda^2}{(H_0 - E_{\eta '})}  H^{(0)} + 
H^{(0)} \frac{\lambda^2}{(H_0 - E_{\eta})}  H^{(0)} \bigg] \eta \,,
\eeq
respectively. Finally, the operators corresponding to the last two graphs in Fig.~\ref{fig5} read:
\beq
\label{5i}
V_{2 \pi}^{\rm 5i} = \eta ' \bigg[ H^{(0)} \frac{\lambda^1}{ \omega }  H^{(0)} 
\frac{\lambda^2}{(\omega_1 + \omega_2)}
H^{(2)} + H^{(2)} \frac{\lambda^2}{( \omega_1 + \omega_2 )} H^{(0)} \frac{\lambda^1}{\omega }  H^{(0)} 
 + H^{(0)} \frac{\lambda^1}{\omega }  H^{(2)}   
\frac{\lambda^1}{\omega } H^{(0)} \bigg] \eta  \,,
\eeq
and 
\beq
\label{5j}
V_{2 \pi}^{\rm 5j} =  \eta '  \bigg[  H^{(2)} \frac{\lambda^2}{(\omega_1 + \omega_2 )}  H^{(0)} + 
H^{(0)} \frac{\lambda^2}{(\omega_1 + \omega_2 )}  H^{(2)} \bigg] \eta \,,
\eeq
where the $\omega$'s refer to the pionic free energy in the isospin limit and 
$H^{(2)}$ denotes the isospin--violating 
$\pi \pi NN$ vertex from the Lagrangian $\mathcal{L}^{(2)}$ proportional to
the LEC $c_5$. 
Performing  a straightforward 
evaluation of the matrix elements of these operators, we obtain the following result for the leading CSB 2PE potential:
\beqa
\label{2PEisosp4}
\label{WC4}
W_C^{(4)} &=& - \frac{g_A^2}{64 \pi F_\pi^4} \left\{ \frac{ 2 g_A^2 \, \delta \bar m  \, M_\pi^3}{4 M_\pi^2 + q^2} 
- \Big( 4 g_A^2 \, \delta \bar m - ( \delta \bar m )^{\rm str} \Big) (2 M_\pi^2 + q^2 ) \, A^\Lambda (q) \right\}\,,  \nn
W_T^{(4)} &=& - \frac{1}{q^2} \, W_S^{(4)} = \frac{g_A^4 \, \delta \bar m}{32 \pi F_\pi^4 } \, A^\Lambda (q )\,,
\eeqa
where the loop function $A^\Lambda (q)$ is given by 
\beq
A^{\Lambda} (q) =  \frac{1}{2 q} \, 
\arctan \frac{q (\Lambda - 2 M_\pi )}{q^2 + 2  \Lambda M_\pi}\,.
\eeq
Here, several comments are in order. First, one should keep in mind that we again only show explicitly the 
non--polynomial terms. Secondly, we found that the planar box graph (e) and
the  ``football'' diagrams (h) and (j) 
lead to vanishing contributions. Notice further that we have also included the
order $\nu =5$ contribution 
from graph (f) in Fig.~\ref{fig6}. The latter is proportional to the LEC $f_2$ or, equivalently, to $(\delta \bar m )^{\rm em}$
and has the same structure as the contribution of diagram (i) in Fig.~\ref{fig5}  which is $\propto c_5$ or, equivalently,
 $\propto (\delta \bar m )^{\rm str}$. It is therefore convenient to combine these contributions and to express them 
in terms of e.g.~$\delta \bar m$ and  $(\delta \bar m )^{\rm str}$. Finally, we would like to emphasize that 
the result given in Eq.~(\ref{WC4}) agrees with previous calculations, see \cite{Niskanen:2001aj,Friar:2003yv}
and \cite{Coon:1995qh} for related older work. 

Consider now the subleading isospin--breaking two--pion--exchange 
potential generated by diagrams shown in Fig.~\ref{fig6}. 
First, charge--symmetry conserving contributions from graphs (a) and (b) due to the pion mass difference can be obtained using 
Eq.~(\ref{2PECIB}) from the corresponding isospin--invariant 2PE potential. We find:
\beq
\label{VT5}
V_T^{(5)} = - \frac{1}{q^2} \, V_S^{(5)} = - \frac{\delta \bar M_\pi^2}{16 \pi F_\pi^4} \,g_A^2 \, c_4\,  A^\Lambda (q)\,. 
\eeq

The contributions of graphs (c) and (d) result in the method of unitary transformation from the operators
\beqa
V_{2 \pi}^{\rm 6c} &=& \frac{1}{2} \eta ' \bigg[ H^{(0)} \frac{\lambda^1}{(H_0 - E_{\eta})}  H^{(0)} 
\frac{\lambda^2}{(H_0 - E_{\eta})}
H^{(1)} + H^{(1)} \frac{\lambda^2}{(H_0 - E_{\eta})} H^{(0)} \frac{\lambda^1}{(H_0 - E_{\eta})}  H^{(0)} \nn
&&{} \mbox{\hskip 0.8 true cm} + H^{(0)} \frac{\lambda^1}{(H_0 - E_{\eta}) }  H^{(1)}   
\frac{\lambda^1}{(H_0 - E_{\eta}) } H^{(0)} \bigg] \eta  + \mbox{h.~c.}\,,
\eeqa
and 
\beq
V_{2 \pi}^{\rm 6d} = \frac{1}{2} \eta '  \bigg[  H^{(1)} \frac{\lambda^2}{(H_0 - E_{\eta '})}  H^{(0)} + 
H^{(1)} \frac{\lambda^2}{(H_0 - E_{\eta})}  H^{(0)} \bigg] \eta  + \mbox{h.~c.} \,,
\eeq
where $H^{(1)}$ refers to $\pi \pi NN$ vertices from $\mathcal{L}^{(1)}$ proportional to $c_i$. 
Further, contributions from graphs (e) and (g) in Fig.~\ref{fig6} can be obtained from $V_{2 \pi}^{\rm 5j}$ in Eq.~(\ref{5j})
by replacing $H^{(0)}$ by $H^{(1)}$ and $H^{(2)}$ by $H^{(3)}$, respectively. The contribution 
of diagram (f) results from $V_{2 \pi}^{\rm 5i}$ in Eq.~(\ref{5i}) with $H^{(2)}$ being replaced by $H^{(3)}$
and has already been included in Eqs.~(\ref{WC4}).
Finally, contributions of graphs (h), (i) and diagram (j) can be obtained by evaluating matrix elements of the operators 
$V_{2 \pi}^{\rm 5e,5f}$ in Eq.~(\ref{5e5f}) and $V_{2 \pi}^{\rm 5g}$ in Eq.~(\ref{5g}), respectively, where one 
of the vertices $H^{(0)}$ is replaced by the isospin--breaking $\pi NN$ vertex $H^{(3)}$ from Eq.~(\ref{lagrI}) and all possible permutations
of the operators  $H^{(3)}$ and  $H^{(0)}$ are taken into account. Notice that in this case one only needs to keep 
the pionic free energy in the corresponding denominators. 
We found that diagrams (d), (g) and (j) lead to vanishing contributions while the subleading CSB potential 
generated by diagrams (c), (e), (h) and (i) in Fig.~\ref{fig6} reads:
\beqa
\label{WC5}
W_C^{(5)} &=& - \frac{1}{96 \pi^2 F_\pi^4} \, L^\Lambda (q) \, \bigg\{ - g_A^2 \, \delta \bar m 
\frac{48 M_\pi^4 (2 c_1 + c_3)}{4 M_\pi^2 + q^2} \nn
&& \mbox{\hskip 3 true cm}  + 4 M_\pi^2 \Big[ g_A^2 \, \delta \bar m \, ( 18 c_1 + 2 c_2 - 3 c_3) 
+ \Big(2 \delta \bar m  - (\delta \bar  m )^{\rm str} \Big) \, ( 6 c_1 - c_2 - 3 c_3 ) \Big] \nn
&& \mbox{\hskip 3 true cm}  + q^2 \Big[ g_A^2 \, \delta \bar m \, ( 5 c_2 - 18 c_3 ) -  
\Big( 2 \delta \bar m - (\delta \bar  m )^{\rm str} \Big)\, ( c_2 + 6 c_3 ) \Big] \bigg\} \,,\nn
W_T^{(5)} &=& - \frac{1}{q^2} \, W_S^{(5)} = - \frac{g_A^2}{16 \pi^2 F_\pi^4} \,  L^\Lambda (q) \,
\Big( \delta \bar m \, c_4 + g_A \, \beta \Big) \,, 
\eeqa
where $\beta =  \epsilon M_\pi^2 \, ( 2 d_{17} - d_{18} - 2 d_{19} )$. 
Notice that we have also included the contribution of diagram (e) in Fig.~\ref{fig6} with the $c_5$--vertex 
being replaced by the $f_2$--vertex from Eqs.~(\ref{lagrI}), which appears formally at order $\nu = 6$.\footnote{The
corresponding contribution $\propto f_1$ does not lead to isospin breaking and is not considered.}
This contribution has precisely the same structure as the one proportional to the LEC $c_5$ with the 
overall strength $2 (\delta \bar m )^{\rm em}$ instead of   $(\delta \bar m )^{\rm str}$.
In Eqs.~(\ref{WC5}) we have expressed $2 (\delta \bar m )^{\rm em} + (\delta \bar m )^{\rm str}$
as $2 \delta \bar m - (\delta \bar m )^{\rm str}$.

To summarize, the charge--symmetry conserving 2PE potential is due to the pion mass difference and includes
a central component at order $\nu = 4$ and tensor and spin--spin components at order $\nu = 5$ 
given in Eqs.~(\ref{VC4}) and (\ref{VT5}), respectively. The CSB 2PE potential has 
all central, tensor and spin--spin components at orders  $\nu = 4$  and  $\nu = 5$, and the corresponding 
expressions are given in Eqs.~(\ref{WC4}) and (\ref{WC5}). 
Further, we stress that our results for the isospin--violating 2PE potential are consistent with 
taking 
\beq
g_A^2 = 4 \pi \, \left( \frac{2 F_\pi}{M_{\pi^\pm}} \right)^2 \, f_c^2
\eeq
in the isospin--conserving 2PE potential. This expression already accounts for the
Goldberger--Treiman discrepancy.
Finally, to the order we are working, the constant $\beta$
can be expressed in terms of the pion--nucleon coupling constants as follows:
\beq
\beta = \frac{f_p^2  - f_n^2}{8 f_c^2} g_A\,.
\eeq

\subsection{Two--pion--exchange potential in coordinate space}
\label{sec:TPEsize}

Let us now take a look at the isospin--violating potential in coordinate space. 
Similar to Eq.~(\ref{2PEmom_gen}) we define:
\beq 
\label{2PEcoord_gen}
V_{2 \pi} = (\tau_1^3 \, \tau_2^3 ) \, \Big[ \tilde V_C + \tilde V_S \, (\vec \sigma_1 \cdot \vec \sigma_2 ) + 
\tilde V_T \, S_{12} \Big] +
(\tau_1^3 +\tau_2^3 ) \, \Big[ \tilde W_C + \tilde W_S \, (\vec \sigma_1 \cdot \vec \sigma_2 ) + 
\tilde W_T \, S_{12} \Big]\,,
\eeq
where $S_{12} = 3 \, \vec \sigma_1 \cdot \hat r \;  \vec \sigma_2 \cdot \hat r - \vec \sigma_1 \cdot \vec \sigma_2$ and 
the functions $\tilde V_C (r), \ldots , \tilde W_T (r)$ depend on the distance $r$. 
Consider now the charge--symmetry--conserving 2PE potentials $\tilde V_i (r)$ which are due to the pion mass difference 
and correspond to $V_C^{(4)}$, $V_T^{(5)}$ and $V_S^{(5)}$ defined in Eqs.~(\ref{VC4}) and (\ref{VT5}).
In order to obtain the $r$--space expressions (at $r \neq 0$), one usually switches to the spectral--function representation 
for the non--polynomial part of the potential which has the form:
\beq
\label{SFR}
V_i (q) = \frac{2}{\pi} \int_{2 M_\pi}^\infty d \mu \, \mu \, \frac{{\rm Im}\big[V_i (- i \mu ) \big]}{\mu^2 + q^2}\,.
\eeq
Here, ${\rm Im}\big[V_i (- i \mu ) \big]$ is the mass spectrum (or the spectral function) entering this representation,
which results from the analytical continuation of the momentum--space functions to $q = 0^+ - i \mu$.
Using this representation one obtains for the functions $\tilde V_{i} (r)$ for $r > 0$:
\beqa
\label{fourc}
\tilde V_{C} (r) &=&  \frac{1}{2 \pi^2 r} \int_{2 M_\pi}^\infty d \mu \, \mu \, 
e^{-\mu r} \, {\rm Im}\big[V_C (- i \mu ) \big] \,, \\
\label{fourt}
\tilde V_{T} (r) &=& -\frac{1}{6 \pi^2 r^3}  \int_{2 M_\pi}^\infty d \mu \, \mu \, 
e^{-\mu r} \, ( 3 + 3 \mu r + \mu^2 r^2 ) \, {\rm Im}\big[V_T (- i \mu )  \big] \,, \\
\label{fours}
\tilde V_{S} (r) &=&  -\frac{1}{6 \pi^2 r}  \int_{2 M_\pi}^\infty d \mu \, \mu \, 
e^{-\mu r} \, \Bigl( \mu^2 \, {\rm Im}\big[V_T (- i \mu ) \big]  - 3 \, {\rm Im}\big[V_S (- i \mu ) \big] \Bigr)\,. 
\eeqa
For the CSC 2PE potential $\propto \delta \bar M_\pi^2$ the spectral--function representation takes a slightly more complicated 
form as compared to Eq.~(\ref{SFR}), namely:
\beq
\label{SFR2}
V_i (q) = \lim_{\epsilon \to 0} \left[ 
\frac{2}{\pi} \int_{2 M_\pi + \epsilon}^\infty d \mu \, \mu \, \frac{{\rm Im}\big[V_i ( - i \mu ) \big]}{\mu^2 + q^2} - 
\frac{8 M_\pi}{\pi} \, \frac{{\rm Im}\big[V_i ( - 2 i M_\pi - i \epsilon ) \big]}{4 M_\pi^2 + q^2}  \right]\,.
\eeq
The second term in this expression accounts for the corresponding shifts of the threshold when pions of different masses 
are exchanged between the nucleons. Notice that both terms in Eq.~(\ref{SFR2}) go to infinity for $\epsilon \to 0$, but their 
sum is, of course, finite in this limit. Although one can, in principle, use this representation on order to obtain $\tilde V_i (r)$,
we choose different strategy by using Eq.~(\ref{2PECIB}) in coordinate space. Substituting the expressions for the leading 
and subleading isospin--invariant 2PE potential in $r$--space \cite{Kaiser:1997mw,Epelbaum:2003gr} in Eq.~(\ref{2PECIB}), we find:
\beqa
\label{Vcoord}
\tilde V_C^{(4)} (r) \bigg|_{\Lambda \to \infty} &=& \frac{\delta \bar M_\pi^2 \, M_\pi}{256 \pi^3 F_\pi^4 \, r^2} 
\left[ 4 g_A^2 (-1 + g_A^2) x K_0 (2 x) + (-1 -6 g_A^2 + g_A^4 (11 + 4 x^2 )) K_1 (2 x) \right] \,,\nn [1ex]
\tilde V_T^{(5)} (r)  &=& \frac{\delta \bar M_\pi^2 \, g_A^2 \, c_4}{192 \pi^2 F_\pi^4}  \, \frac{e^{-2 x}}{r^4} (4 + x (5 + 2 x))
-  \frac{\delta \bar M_\pi^2 \, g_A^2 \, c_4}{384 \pi^2 F_\pi^4}  \, \frac{e^{-y}}{r^4} (8 + y (5 + y)) \,, \nn [1ex]
\tilde V_S^{(5)} (r)  &=& - \frac{\delta \bar M_\pi^2 \, g_A^2 \, c_4}{96 \pi^2 F_\pi^4}  \, \frac{e^{-2 x}}{r^4} (1 + 2 x (1 + x))
+  \frac{\delta \bar M_\pi^2 \, g_A^2 \, c_4}{192 \pi^2 F_\pi^4}  \, \frac{e^{-y}}{r^4} (2 + y (2 + y))\,,
\eeqa
where $x = M_\pi r$, $y = \Lambda r$ and $K_i$ are the modified Bessel functions. 
In the case of $\tilde V_C^{(4)}$, we could only perform the integral in 
Eq.~(\ref{fourc}) analytically for $\Lambda \to \infty$
which corresponds to the DR result. 

\begin{figure*}[tb]
\vspace{0.5cm}
\centerline{
\psfig{file=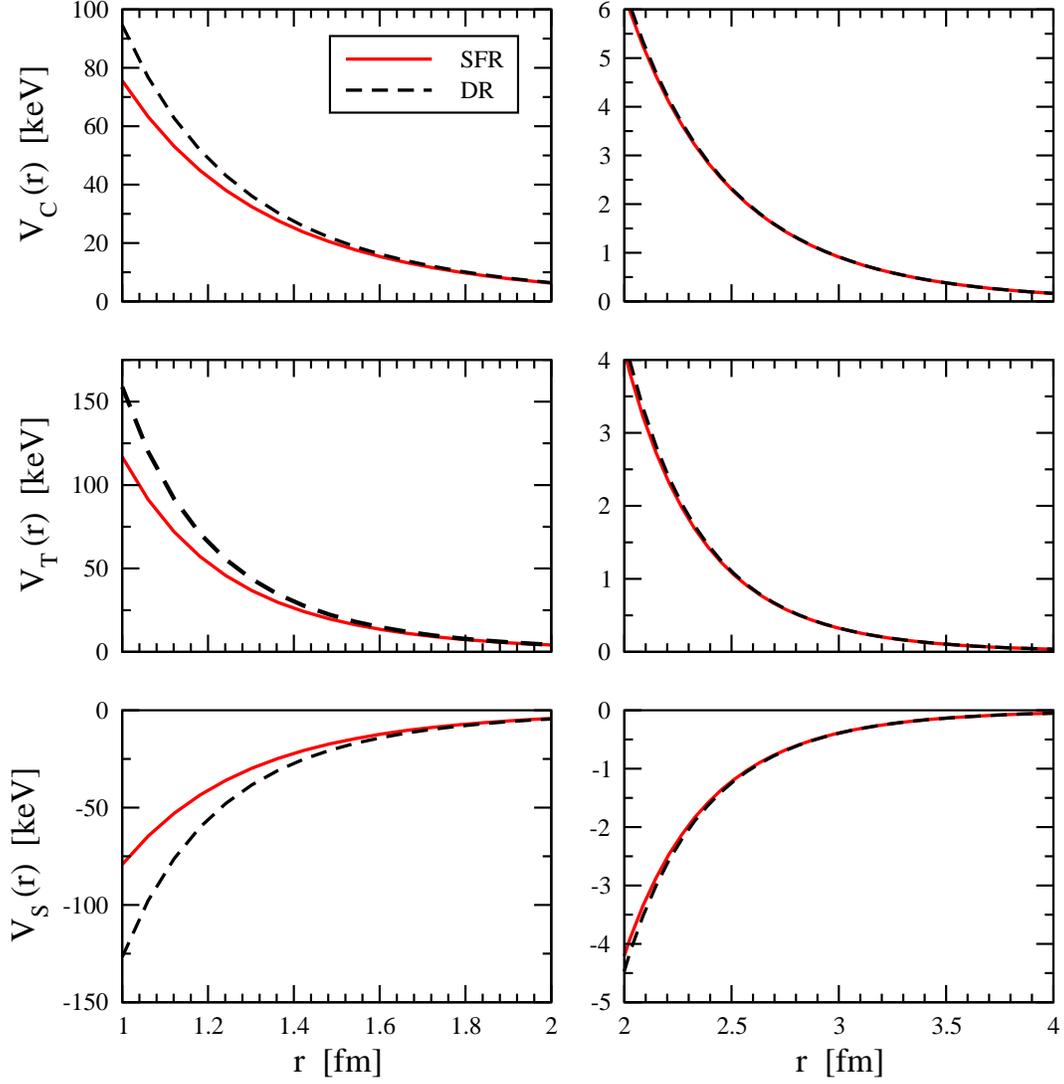,width=15cm}
}
\vspace{0.3cm}
\centerline{\parbox{14cm}{
\caption[fig7]{\label{fig7} Central (upper row), tensor (middle row) and spin--spin (bottom row) components of the  
charge--symmetry--conserving (CSC) 2PE potential in coordinate space,
utilizing DR (dashed) and SFR (solid lines). The SFR result corresponds to $\Lambda = 700$ MeV.
Left/right panels: Distances from $1\ldots 2$/$2\ldots 4$~fm.
}}}
\vspace{0.3cm}
\end{figure*}

To obtain the CSB 2PE at orders $\nu =4$ and $\nu =5$ in $r$--space, it is convenient to use  
Eqs.~(\ref{fourc})--(\ref{fours}) (with $V_i$ being replaced by $W_i$). In that case, the spectral--function representation
in Eq.~(\ref{SFR}) is valid. Using Eqs.~(\ref{WC4}), (\ref{WC5}) we find:
\beqa
\tilde W_C^{(4)} (r) &=& \frac{g_A^2}{256 \pi^2 F_\pi^4} \, \frac{e^{-2 x}} {r^4} \, 
\left( ( \delta \bar m )^{\rm str}  (1 + x )^2 - 2 \delta \bar m \, g_A^2 (2 + x (4 + x (2 + x))) \right)\nn
&& {} - \frac{g_A^2}{512 \pi^2 F_\pi^4} \, \frac{e^{-y}} {r^4} \, (( \delta \bar m )^{\rm str}  - 4 \delta \bar m \, g_A^2 ) 
\left(2 - 2 x^2 + y (2 + y) \right) \,,\nn [1ex]
\tilde W_T^{(4)} (r) &=& - \frac{\delta \bar m \, g_A^4}{384 \pi^2 F_\pi^4} \, \frac{e^{- 2 x}}{r^4} 
\left(4 + x (5 + 2 x) \right) 
+\frac{\delta \bar m \, g_A^4}{768 \pi^2 F_\pi^4} \, \frac{e^{- y}}{r^4} \left(8 + y (5 + y) \right) \,,\nn [1ex]
\tilde W_S^{(4)} (r) &=&  \frac{\delta \bar m \, g_A^4}{192 \pi^2 F_\pi^4} \, \frac{e^{- 2 x}}{r^4} 
\left(1 + 2 x + 2 x^2 ) \right) 
-\frac{\delta \bar m \, g_A^4}{384 \pi^2 F_\pi^4} \, \frac{e^{- y}}{r^4} \left(2 + 2 y + y^2) \right) \,,
\eeqa
and 
\beqa
\tilde W_C^{(5)} (r) \bigg|_{\Lambda \to \infty} &=& \frac{M_\pi}{32 \pi^3 F_\pi^4 \, r^4} \bigg[ \Big(
g_A^2 \, \delta \bar m \, x ( - 5 c_2 + 18 c_3 + 4 (2 c_1 + c_3) x^2 ) \nn
&&{} + \big( 2 \delta \bar m - (\delta \bar m )^{\rm str} \big) \, x \, (c_2 + 6 c_3 ) 
\Big) K_0 (2 x) \nn
&&{} + \Big( g_A^2 \, \delta \bar m \, (-5 c_2 + 18 c_3 + 2 (6 c_1 - c_2 + 5 c_3 ) x^2 ) \nn
&&{} + \big( 2 \delta \bar m - (\delta \bar m )^{\rm str} \big)  \, (c_2 + 6 c_3 + 2 (2 c_1 + c_3 ) x^2) \Big) K_1 (2 x) \bigg] \,, \nn [1ex]
\tilde W_T^{(5)} (r) \bigg|_{\Lambda \to \infty} &=& - \frac{g_A^2 \, M_\pi}{96 \pi^3 F_\pi^4 \, r^4} 
(c_4 \, \delta \bar m + \beta \, g_A ) \Big( 12 x \, K_0 (2 x) + (15 + 4 x^2) K_1 (2 x) \Big)\,, \nn [1ex]
\tilde W_S^{(5)} (r) \bigg|_{\Lambda \to \infty} &=&  \frac{g_A^2 \, M_\pi}{24 \pi^3 F_\pi^4 \, r^4} 
(c_4 \, \delta \bar m + \beta \, g_A ) \Big( 3 x \, K_0 (2 x) + (3 + 2 x^2) K_1 (2 x) \Big) \,.
\eeqa
Again, at order $\nu = 5$ we could only perform integrals in Eqs.~(\ref{fourc})--(\ref{fours}) analytically for $\Lambda \to \infty$.

\begin{figure*}[tb]
\vspace{0.5cm}
\centerline{
\psfig{file=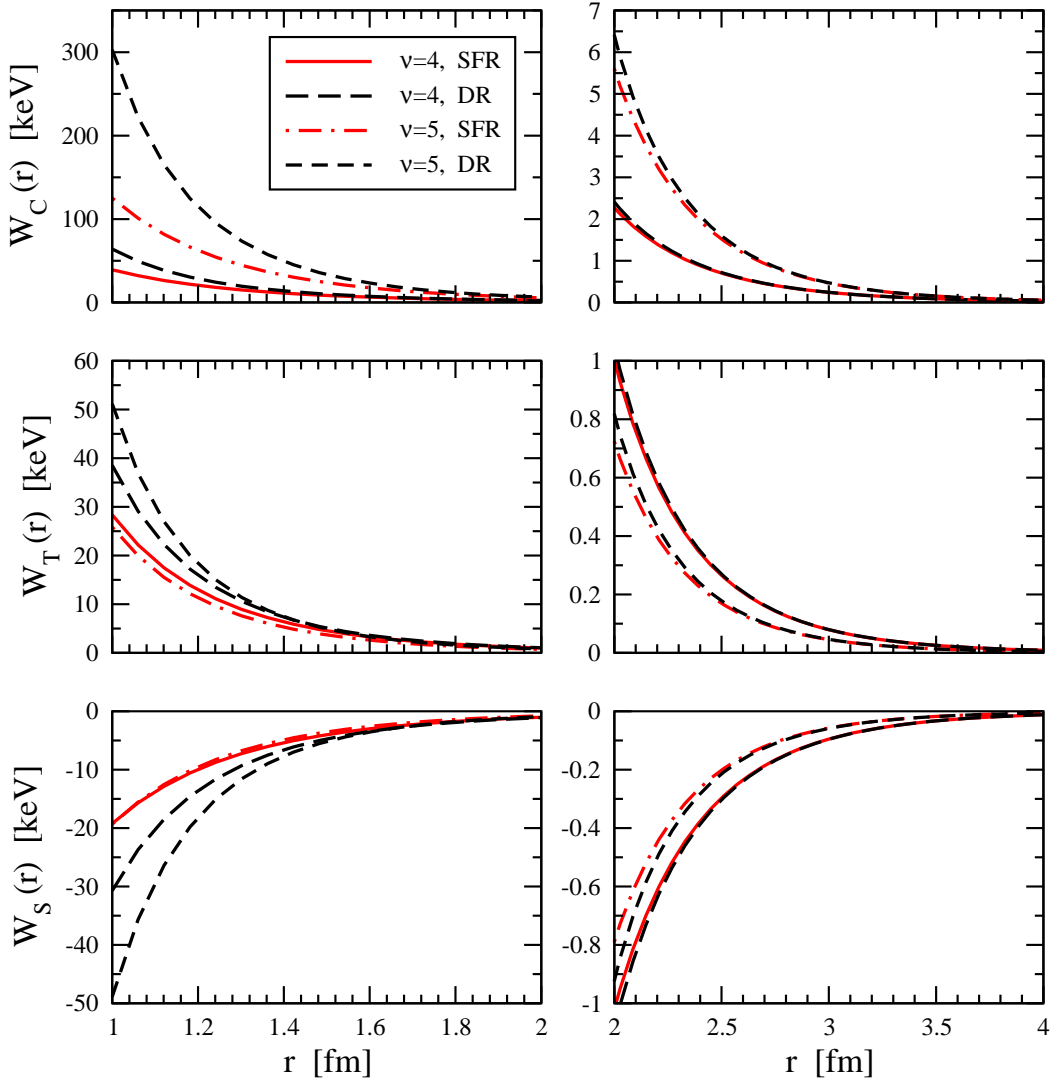,width=15cm}
}
\vspace{0.3cm}
\centerline{\parbox{14cm}{
\caption[fig8]{\label{fig8} Central (upper row), tensor (middle row) and spin--spin (bottom row) components of the  
charge--symmetry--breaking 2PE potential in coordinate space.
Long-/short-dashed lines: DR at  $\nu = 4$ and $\nu = 5$; 
Solid/dot-dashed lines: SFR at  $\nu = 4$ and $\nu = 5$.
The SFR results correspond  to $\Lambda = 700$ MeV.
}}}
\vspace{0.3cm}
\end{figure*}

\begin{figure*}[tb]
\vspace{0.5cm}
\centerline{
\psfig{file=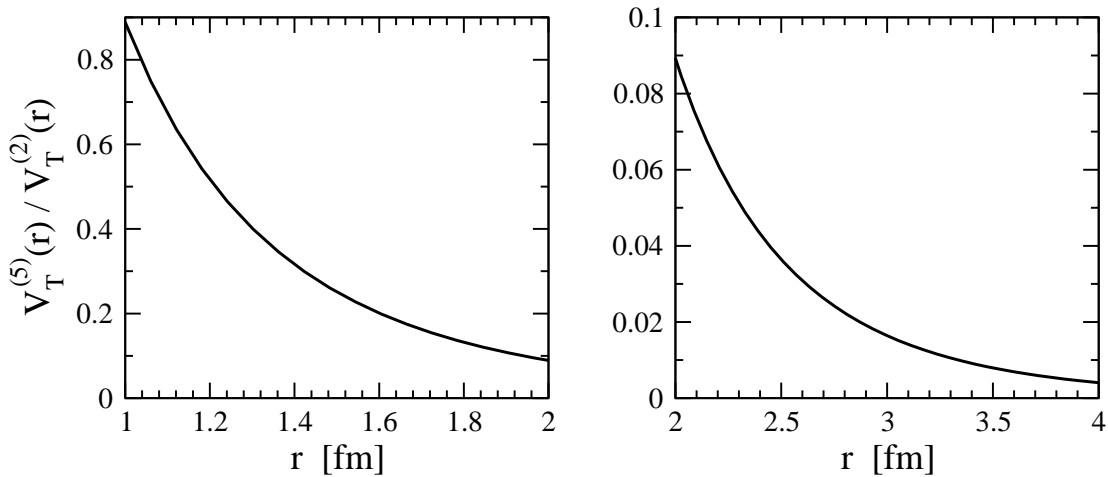,width=15cm}
}
\vspace{0.3cm}
\centerline{\parbox{14cm}{
\caption[fig10]{\label{fig10} Ratio of the CSC tensor $2\pi$--exchange  
(using SFR with $\Lambda = 700$ MeV) 
and $1\pi$--exchange potentials as a function of $r$. 
}}}
\vspace{0.3cm}
\end{figure*}

It is now interesting to compare the strength of the corresponding $r$--space potentials. Here and in what follows, 
we adopt the same values for the LECs $c_i$ as in 
our work \cite{Epelbaum:2004fk}: $c_1 = -0.81$ GeV$^{-1}$, $c_2 = 3.28$ GeV$^{-1}$, $c_3 = -3.40$ GeV$^{-1}$ and $c_4 = -3.40$ GeV$^{-1}$. 
Further, $g_A=1.27$, 
$M_\pi = 138.03$ MeV and $F_\pi = 92.4$ MeV. For the strong nucleon mass shift $(\delta m )^{\rm str}$,
we use the value given in Eq.~(\ref{m_shift2}). Finally, in our numerical estimations we set $\beta = 0$ since the 
value of this LEC is not known at present.  
Notice, however, that a 1\% relative deviation between $f_p$ and $f_n$  leads to $g_A  \beta \sim \delta \bar m \, c_4$,
so that the strength of the resulting CSB potential is comparable to the one of the CSB potential $\propto c_4$.
The CSC and CSB 2PE potentials are plotted in Figs.~\ref{fig7} and \ref{fig8}, respectively, for two choices of the cut--off $\Lambda$
in the SFR: $\Lambda = 700$ MeV and $\Lambda = \infty$ which is equivalent to DR. The results for $\tilde V_C^{(4)} (r)$  and 
$\tilde W_i^{(5)} (r)$ for $\Lambda = 700$ MeV are obtained numerically. First of all, we notice that all CSC 2PE contributions
have similar strength $\sim 100$ keV at $r = 1$ fm and $\sim 4 - 6$ keV at $r = 2$ fm. Although the central potential 
$V_C^{(4)} (r)$ is formally dominant, the subleading contributions $V_{T,S}^{(5)} (r)$ are numerically large due to the large 
value of the LEC $c_4$. To get further insights into the importance of the CSC 2PE potential, it is instructive 
to compare its strength with the strength of the corresponding 1PE potential resulting due to the pion mass 
difference, which provides the dominant isospin--violating contribution to the NN force at order $\nu = 2$. It has the following 
form in momentum space:
\beq
V_T^{(2)} (q) = - \left( \frac{g_A}{2 F_\pi} \right)^2 \delta \bar M_\pi^2 \, \frac{(\vec \sigma_1 \cdot \vec q \,)
(\vec \sigma_2 \cdot \vec q \,)}{(q^2 + M_\pi^2 )^2}\,.
\eeq
The corresponding $r$--space expressions read (for $r > 0$):
\beq
\tilde V_T^{(2)} (r) = \frac{g_A^2 \, \delta \bar M_\pi^2}{96 \pi F_\pi^2} \, \frac{e^{- x}}{r} (1 + x)\,, \quad \quad \quad
\tilde V_S^{(2)} (r) = -\frac{g_A^2 \, \delta \bar M_\pi^2}{96 \pi F_\pi^2} \, \frac{e^{- x}}{r} (2 - x)\,.
\eeq
In Fig.~\ref{fig10} we have plotted the ratio $\tilde V_T^{(5)} / \tilde V_T^{(2)}$ as a function of $r$. The 2PE contribution is
significant for $r \lesssim 2$ fm. At larger distances it becomes negligible (less than 1\% for $r \gtrsim 3.5$ fm) compared to the 
1PE contribution due to its shorter range. 

\begin{figure*}[tb]
\vspace{0.5cm}
\centerline{
\psfig{file=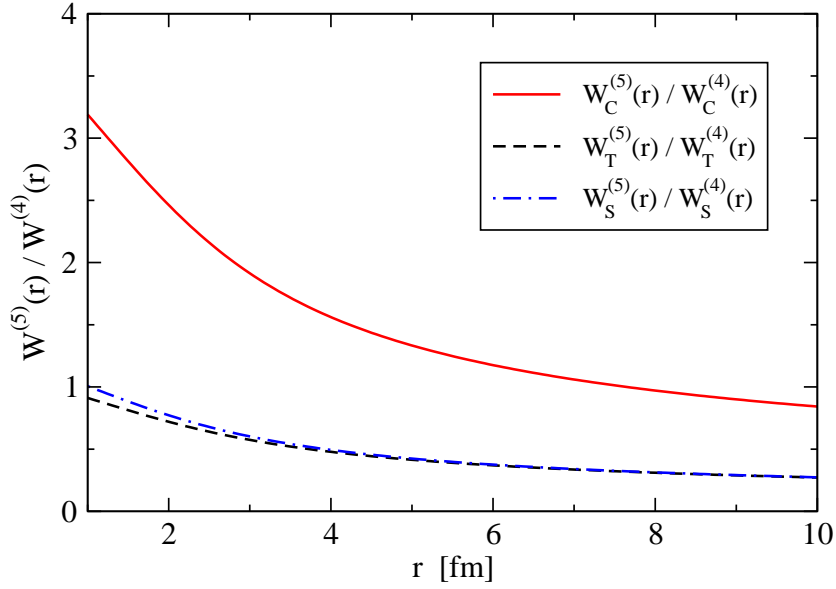,width=12cm}
}
\vspace{0.3cm}
\centerline{\parbox{14cm}{
\caption[fig9]{\label{fig9} Ratio of the subleading ($\nu =5$) and leading ($\nu =4$) CSB 2PE potential 
as a function of $r$. Solid/dashed/dot-dashed line: central/tensor/spin-spin component.
All results correspond to SFR with $\Lambda = 700$ MeV.
}}}
\vspace{0.3cm}
\end{figure*}

Let us now switch to the CSB contributions displayed in Fig.~\ref{fig8}.
Similar to the CSC potential, the subleading terms $\propto c_i$ are numerically enhanced due to large values of these LECs.
The strongest contribution
is given by the subleading central potential which reaches $\sim 150 \, - \, 300$ keV  at $r = 1$ fm and $\sim 6$ keV at $r = 2$ fm
and is dominated by terms $\propto c_3$. This is similar to the isospin--symmetric case, where the subleading central 2PE 
potential is known to be very strong. 
To enable a more detailed comparison between the leading and subleading contributions to the CSB 
2PE potential, we plot  in Fig.~\ref{fig9} the corresponding ratios.
While the subleading central component is significantly stronger than the
leading one in a wide range of distances $r$, 
the tensor and spin--spin components are of the same size as the leading ones only for $r  \lesssim 2$ fm. One should,
however, keep in mind that our results obtained within the heavy baryon formalism become formally invalid at very large distances.  
This problem with the heavy baryon formalism has been first observed in the single--nucleon sector and can be dealt with 
using e.g. the Lorentz invariant scheme proposed by Becher and Leutwyler \cite{Becher:1999he}, see also \cite{Fuchs:2003qc}. 
It is clear, however, that the NN interaction due to two--pion exchange becomes very weak at 
large distances, so that the problem with the formal inconsistency of the heavy baryon   
approach is expected to have little relevance for practical applications, see \cite{Higa:2004cr} for more details. 

Finally, we would like to emphasize that the SFR results  for CSC and the leading CSB contributions 
are rather close to the ones obtained in DR
even for $1\mbox{ fm} < r < 1.5\mbox{ fm}$, see Fig.~\ref{fig7}. Stated differently, the corresponding potentials seem to be 
strongly dominated by long--range components in 2$\pi$--exchange. Deviations between the SFR and DR results at short distances  
are significantly larger for CSB contributions at order $\nu = 5$. This is consistent with the fact that new CSB short--range 
terms with two derivatives start to contribute at this order. 
One expects that such new counterterms will largely  
remove the dependence of observables on the cut--off in the spectral--function representation.

\subsection{Contact terms}
\label{sec:NNcont}

We now discuss short--range isospin--breaking interactions. 
The leading contact terms in the NN potential at $\nu = 3$ are proportional to the quark mass difference and violate charge symmetry:
\beq
\label{sr1}
V_{\rm cont}^{(3)} = \beta_{3, 1} \, (\tau_1^3 + \tau_2^3) + \beta_{3, 2} \; \vec \sigma_1 \cdot \vec \sigma_2 \, 
(\tau_1^3 + \tau_2^3) \,,
\eeq
where $\beta_{3, i} \sim \epsilon M_\pi^2/( F_\pi^2 \Lambda^2)$ are  constants. 
We remind the reader that the subscripts of the spin and isospin matrices refer to nucleon labels while 
the superscripts denote the corresponding vector indices. 
Both terms in Eq.~(\ref{sr1})
lead to the same structure in the potential when anti-symmetrization with respect to the nucleons is performed. Consequently, it is sufficient to 
keep only one term.\footnote{Equivalently, one can apply a Fierz transformation in the corresponding Lagrangian in order to eliminate 
redundant terms.}  At order $\nu = 4$ one has to take into account isospin--violating short--range interactions without 
derivatives of electromagnetic origin. In addition to the terms which have the same structure as the ones in Eq.~(\ref{sr1})
and thus only provide order $\sim e^2/(4 \pi F_\pi )^2$ shifts of $\beta_{3, i}$, new CSC interactions appear:
\beq
\label{sr2}
V_{\rm cont}^{(4)} = \beta_{4, 1} \, \tau_1^3 \tau_2^3 + \beta_{4, 2} \; \vec \sigma_1 \cdot \vec \sigma_2  \, \tau_1^3 \tau_2^3
\eeq
Again, one of these two terms can be eliminated performing anti--symmetrization of the potential. 
Finally, at order $\nu = 5$ one needs to include CSB terms with two derivatives which are proportional 
to the quark mass difference. In the two--nucleon CMS these terms read: 
\beqa
\label{sr3}
V_{\rm cont}^{(5)} &=& (\tau_1^3 + \tau_2^3) \bigg[
\beta_{5, 1} \, \vec q \,^2 + \beta_{5, 2} \, \vec k \, ^2 + ( \beta_{5, 3} \, \vec q \,^2 + \beta_{5, 4} \, \vec k \, ^2 ) 
(\vec \sigma_1 \cdot \vec \sigma_2 ) + \frac{i}{2}   \beta_{5, 5} \, (\vec \sigma_1 + \vec \sigma_2 ) \cdot ( \vec k \times \vec q \, )\nn
&& {}  + \beta_{5, 6} \,(\vec \sigma_1 \cdot \vec q \, ) (\vec \sigma_2 \cdot \vec q \, ) 
+ \beta_{5, 7} \,(\vec \sigma_1 \cdot \vec k \, ) (\vec \sigma_2 \cdot \vec k \, ) \bigg]
+ i \beta_{5, 8} \,(\tau_1^3 - \tau_2^3) \, \vec k \times \vec q \cdot ( \vec \sigma_1 - \vec \sigma_2 ) \\
&& {}  + i \beta_{5, 9} \, [\fet \tau_1 \times  \fet \tau_2 ]^3 \, \vec k \times \vec q \cdot [ \vec \sigma_1 \times  \vec \sigma_2 ] 
+ i \beta_{5, 10} \, [\fet \tau_1 \times  \fet \tau_2 ]^3 \, \Big((\vec \sigma_1 \cdot \vec q \, ) (\vec \sigma_2 \cdot \vec k \, ) -
(\vec \sigma_1 \cdot \vec k \, ) (\vec \sigma_2 \cdot \vec q \, ) \Big)\,,
\nonumber
\eeqa
where  $\beta_{5, i} \sim \epsilon M_\pi^2/(F_\pi^2 \Lambda^4)$ are further constants and 
$\vec k =  ( \vec p + \vec p\,')/2$. 
Performing anti--symmetrization of this potential, it is easy to see that half of the terms 
proportional to $\beta_{5, 1}$, $\beta_{5, 2}$, $\beta_{5, 3}$, $\beta_{5, 4}$,
$\beta_{5, 6}$ and $\beta_{5, 7}$ are redundant. Similarly, terms proportional to 
$\beta_{5, 8}$, $\beta_{5, 9}$ and $\beta_{5, 10}$, which lead to mixing between the $T=1$ and $T=0$ states,
generate the same structure when the potential is anti--symmetrized. We are, therefore, left with five independent 
short--range terms at order $\nu = 5$ 
(for example, one can take terms proportional to $\beta_{5, 1}$, $\beta_{5, 3}$, $\beta_{5, 5}$, $\beta_{5, 6}$
and $\beta_{5, 8}$). 

The isospin--breaking short--range terms up to order $\nu = 5$ feed into the matrix--elements of the S-- and P--waves  
in the following way:
\beqa
\label{VC}
\langle ^1S_0, \; pp \, | V_{\rm cont}| ^1S_0, \; pp \, \rangle&=& \tilde \beta_{1S0}^{pp} + \beta_{1S0}\, ( p^2 + p '^2)~,\nn
\langle ^1S_0, \, nn \, | V_{\rm cont}| ^1S_0, \, nn \, \rangle&=& \tilde \beta_{1S0}^{nn} - \beta_{1S0}\, ( p^2 + p '^2)~,\nn
\langle ^3P_0, \; pp \, | V_{\rm cont}| ^3P_0, \; pp \, \rangle&=& 
- \langle ^3P_0, \; nn \, | V_{\rm cont}| ^3P_0, \; nn \, \rangle = \beta_{3P0}\,  \,p \, p' ~,\nn
\langle ^3P_1, \; pp \, | V_{\rm cont}| ^3P_1, \; pp \, \rangle&=& 
- \langle ^3P_1, \; nn \, | V_{\rm cont}| ^3P_1, \; nn \, \rangle = \beta_{3P1}\,  \,p \, p' ~,\nn
\langle ^3P_2, \; pp \, | V_{\rm cont}| ^3P_2, \; pp \, \rangle&=& 
- \langle ^3P_2, \; nn \, | V_{\rm cont}| ^3P_2, \; nn \, \rangle = \beta_{3P2}\,  \,p \, p' ~,\nn
\langle ^1P_1, \, np \, | V_{\rm cont}| ^3P_1, \, np \, \rangle&=& \beta_{1P1-3P1}\,  \,p \, p'  ~,
\eeqa
where the new LECs $\tilde \beta_{1S0}^{pp}$,  $\tilde \beta_{1S0}^{nn}$,  $\beta_{1S0}$, 
 $\beta_{3P0}$,  $\beta_{3P1}$ and  $\beta_{1P1-3P1}$ can be expressed in terms of linear combinations 
of the LECs $\beta_{3, i}$, $\beta_{4, i}$ and $\beta_{5, i}$. Notice that we have adopted here 
the convention according to which the {\it np} matrix elements (with exception
of the last term in Eq.~(\ref{VC}))
do not change by switching off isospin--violating contact terms.

As pointed out in \cite{Walzl:2000cx}, one should, in principle, also take into account isospin--violating CSC contact terms
associated with the contributions to the 2PE potential $\propto  M_\pi^2$. These terms arise since the derivative in 
Eq.~(\ref{2PECIB}) has to be applied not only to the non--polynomial part of $V_{2 \pi}^1 (M_\pi )$, but also 
to the corresponding counterterms which depend on $M_\pi$. The resulting CSC contact terms have fixed coefficients 
(in terms of $g_A$ and $F_\pi$) and are of the order $\sim \delta \bar M_\pi^2/M_\pi^2$ instead of expected
$\sim \delta \bar M_\pi^2/\Lambda^2$ compared to the corresponding isospin--invariant terms. 
We have found that these contact interactions cancel against the ones which result from taking the derivative 
of the non--polynomial part of  $V_{2 \pi}^1 (M_\pi )$. Thus, Eqs.~(\ref{VC4}) and (\ref{VT5})
give the complete result for the corresponding CSC 2PE potential, and no additional contact terms need to be taken into account.
Finally, we stress that no contact interactions depending on the total two--nucleon momentum $\vec P$ (similar to Eqs.~(\ref{OPEP2}),
(\ref{OPEP1})) appear up to the order $\nu = 5$.

\section{Consistency of the 2N and 3N forces}
\label{sec:2N3N}

\begin{figure*}[tb]
\vspace{0.5cm}
\centerline{
\psfrag{x11}{\raisebox{-0.0cm}{\hskip 0.1 true cm  (a)}}
\psfrag{x22}{\raisebox{-0.0cm}{\hskip 0.1 true cm  (b)}}
\psfig{file=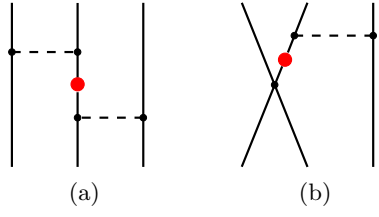,width=5cm}
}
\vspace{0.3cm}
\centerline{\parbox{14cm}{
\caption[fig11]{\label{fig11} Isospin violating 3NF diagrams which include reducible topologies.
For  notation, see Figs.~\ref{fig1}, \ref{fig2}. 
}}}
\vspace{0.3cm}
\end{figure*}

It is well known that two-- and three--nucleon forces   
have to be consistent with each other. In \cite{Epelbaum:2004xf} and
\cite{Epelbaum:2004qe}
we have derived isospin--violating 3NF at orders $\nu = 4$ and $\nu = 5$. It has 1PE, 2PE and contact pieces and 
results from insertions of pion and nucleon mass differences as well as the $c_5$-- and $f_{1,2}$--vertices in Eq.~(\ref{lagrI}). 
Two 3NF diagrams out of seven shown in Fig.~1 of Ref.~\cite{Epelbaum:2004qe} 
appear to be of a special interest due to the fact that they include reducible topologies, i.e.~time--ordered graphs with 
purely nucleonic intermediate states. These two diagrams are depicted in Fig.~\ref{fig11} and lead to 
the following contributions to the 3NF at order $\nu = 4$:
\beqa
\label{3NFisosp1}
V^{\rm 11a} &=& \sum_{i \not= j \not= k} \,2 \delta \bar m  \,  \left(
  \frac{g_A}{2 F_\pi} \right)^4 \frac{( \vec \sigma_i \cdot \vec q_{i}  ) 
(\vec \sigma_j \cdot \vec q_j  )}{(\vec q_i{} ^2 + M_{\pi}^2 )^2 ( \vec
q_j{} ^2 + M_{\pi}^2)} \bigg\{ [\vec q_i \times \vec q_j ] \cdot \vec \sigma_k  \, [ \fet \tau_i \times \fet \tau_j ]^3
\nn
&& \mbox{\hspace{5cm}}
+ \vec q_i \cdot \vec q_j \left[ (\fet \tau_i \cdot \fet \tau_k ) \tau_j^3  -
(\fet \tau_i \cdot \fet \tau_j ) \tau_k^3 \right] \bigg\}\,, \nn
V^{\rm 11b} &=& \sum_{i \not= j \not= k} 2 \, \delta \bar m \, C_T   \left( \frac{g_A}{2 F_\pi} \right)^2
\frac{\vec \sigma_i \cdot \vec q_i}{(\vec q_i {} ^2 + M_{\pi}^2)^2} \, [ \fet \tau_k \times \fet \tau_i ]^3 \; 
[ \vec \sigma_j \times \vec \sigma_k ] \cdot \vec q_i \,, 
\eeqa
where $i$, $j$ and $k$ are nucleon labels and $\vec q_i = \vec p_i \, ' - \vec
  p_i$. Further, $C_T$ is one of the two leading order four--nucleon LECs \cite{Weinberg:1991um}. 
Notice that these nonvanishing contributions are in strong contrast to the corresponding isospin--invariant 3NF 
forces $\propto g_A^4$ and $\propto g_A^2 \, C_i$
as well as isospin--breaking ones $\propto \delta \bar M_\pi^2 \, g_A^4$ and $\propto \delta \bar M_\pi^2 \, g_A^2 \, C_i$
which are known to vanish, see e.g.~\cite{Yang:1986pk,Eden:1996ey,vanKolck:1994yi}. 
It is, therefore, instructive to look at the origin of these particular 3NFs in more detail. 
To that aim we evaluate the 3N scattering amplitude $\propto \delta \bar m \,  g_A^4$ and $\propto \delta \bar m \, g_A^2 \, C_T$ and 
compare the result with the iterated 2N potential. 
We will demonstrate that the iterated 2NF + 3NF  given in 
Eq.~(\ref{3NFisosp1}) reproduces correctly the 3N scattering amplitude. This is an excellent and rather non--trivial check 
of our results. 

\begin{figure*}[tb]
\vspace{0.5cm}
\centerline{
\psfrag{x11}{\raisebox{-0.0cm}{\hskip 0.1 true cm  (a)}}
\psfrag{x22}{\raisebox{-0.0cm}{\hskip 0.1 true cm  (b)}}
\psfrag{x33}{\raisebox{-0.0cm}{\hskip 0.1 true cm  (c)}}
\psfrag{x44}{\raisebox{-0.0cm}{\hskip 0.1 true cm  (d)}}
\psfig{file=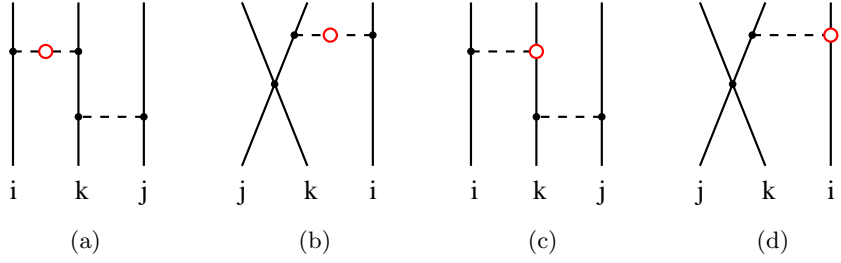,width=11.5cm}
}
\vspace{0.3cm}
\centerline{\parbox{14cm}{
\caption[fig12]{\label{fig12} Feynman diagrams contributing to the CSB 3N scattering amplitude. 
Graphs resulting from the interchange of the nucleon lines and/or application of the time reversal
operation are not shown. Empty circles denote insertions of the $\delta \bar m$--vertices from Eq.~(\ref{lagr_new}).  
}}}
\vspace{0.3cm}
\end{figure*}

The most convenient way to evaluate the scattering amplitude is using the Feynman graph technique. 
In the following, we will use the method suggested in Ref.~\cite{Friar:2004ca}, in which the proton--to--neutron 
mass difference is removed from the Lagrangian by an appropriate field redefinition in favor of new isospin--violating 
terms $\propto \delta \bar m$ which are easier to handle in practical applications. Only two of such new isospin--breaking 
terms may lead to contributions $\propto \delta \bar m \,  g_A^4$ and $\propto \delta \bar m \, g_A^2 \, C_T$ in the amplitude:
\beq
\label{lagr_new}
\mathcal{L '} = \delta \bar m (\fet \pi \times \dot{\fet \pi } )_3  - \frac{g_A}{4 F_\pi} \frac{\delta \bar m}{m} 
N^\dagger \{ \vec \sigma \cdot \vec p, \; ( \fet \tau \times \fet \pi )_3 \} N~.
\eeq
Notice that in contrast to nuclear forces, the on--shell scattering amplitude we are interested in is unique and 
does not depend on field redefinitions. The relevant Feynman diagrams are shown in Fig.~\ref{fig12}. 
Notice that Feynman graphs resulting from diagrams (a) and (b) with one of the $g_A$--vertices being replaced by 
the vertex corresponding to the second term in Eq.~(\ref{lagr}) lead to contributions $\propto 1/m$
which are irrelevant for our discussion. The contribution from graph (a) is given by:
\beqa
\label{pr0}
T^{12a} &=& i \left( \frac{g_A}{2 F_\pi} \right)^4 \, \frac{i^2}{[q_i{}^2 - M_\pi^2 + i \epsilon ]^2}\, 
\frac{i \, 2 m \Lambda_+}{[\tilde p_k{}^2 - m^2 + i \epsilon ]} \,
\frac{i}{[q_j{}^2 - M_\pi^2 + i \epsilon ]} \nn
&& {} \mbox{\hskip 1.8 true cm} \times (\vec \sigma_i \cdot \vec q_i )
 (-\vec \sigma_k \cdot \vec q_i )  (\vec \sigma_k \cdot \vec q_j )  (-\vec \sigma_j \cdot \vec q_j )
\, (-2 \delta \bar m) \, (q_i)_0 \, \epsilon_{ab3} \,  \tau_i^b \tau_k^a \,  (\fet \tau_k \cdot \fet \tau_j )\,, 
\eeqa
where we use the relativistic nucleon propagator. Here, $\Lambda_+$ is the 
projection matrix onto positive--energy
states and $\tilde p_k = p_k ' + q_i = p_k' + p_i '  - p_i = 
\Big(\sqrt{{\vec p_k {}'}^2  + m^2} + 
\sqrt{{\vec p_i {}'} ^2  + m^2} - \sqrt{\vec p_i {}^2  + m^2}, \; \vec p_k\, '  + \vec q_i \Big)$
is the four--momentum of the nucleon $k$ in the intermediate state.
Notice further that the amplitude is multiplied by $i$ in order to match with the standard 
normalization of the nonrelativistic $T$--matrix. The consistency of this procedure can be verified 
e.g.~by calculating the static isospin--invariant 1PE potential. Using the relation:
\beq
\frac{i \, 2 m \Lambda_+}{[\tilde p_k{}^2 - m^2 + i \epsilon ]} = \frac{i}{\delta T + i \epsilon } 
+ \mathcal{O} \left( \frac{1}{m} \right)\,,
\eeq
where $\delta T = ( {\vec p_i {}'}^2  +  {\vec p_k {}'}^2 -  \vec p_i {}^2  -  \vec{\tilde p_k} {}^2)/2m$,
we can express $T^{12a}$ as (modulo $1/m^2$--corrections):
\beq
\label{Tprom}
T^{12a} = - i \, 2 \, \delta \bar m \, v_{ik}^2 \, [ \fet \tau_i \times \fet \tau_k]^3 \, \left( \frac{{\vec p_i {}'}^2}{2 m}
-  \frac{\vec p_i {}^2}{2 m} \right) \, \frac{1}{\delta  T + i \epsilon }  \, v_{kj}^1 \, (\fet \tau_k \cdot \fet \tau_j )
\,,
\eeq
where 
\beq
v_{ik}^n = - \left( \frac{g_A}{2 F_\pi} \right)^2 \;
\frac{(\vec \sigma_i \cdot \vec q_i\, )(\vec \sigma_k \cdot \vec q_i \,)}{(\vec{q_i} ^2 + M_\pi^2 )^n} \,.
\eeq
We now use the equality
\beq
\left( \frac{{\vec p_i {}'}^2}{2 m} -  \frac{\vec p_i {}^2}{2 m} \right) =
\frac{1}{2} \left( \frac{{\vec p_i {}'}^2}{2 m} -  \frac{\vec p_i {}^2}{2 m}  
- \frac{{\vec p_k {}'}^2}{2 m} +  \frac{\vec{\tilde p_k} {}^2}{2 m} + \delta T \right)\,,
\eeq
to rewrite Eq.~(\ref{Tprom}) in the form:
\beqa 
\label{Tfin}
T^{12a} &=& i \, \frac{\delta \bar m}{2 m}  \, v_{ik}^2 \, [ \fet \tau_i \times \fet \tau_k]^3 \,
\Big( \vec p_i {}^2 - \vec{\tilde p_k} {}^2  -{\vec p_i {}'}^2 + {\vec p_k {}'}^2 \Big)
\, \frac{1}{\delta  T + i \epsilon }  \, v_{kj}^1 \, (\fet \tau_k \cdot \fet \tau_j ) \nn
&& {}  -i \, \delta \bar m  \, v_{ik}^2 \, [ \fet \tau_i \times \fet \tau_k]^3 \, 
 v_{kj}^1 \, (\fet \tau_k \cdot \fet \tau_j )\,.
\eeqa
The first term in the above equation can be identified with the iteration of the isospin--violating 
1PE potential in Eq.~(\ref{OPEP1}) between the nucleons $i$ and $k$ and the isospin--invariant static 1PE potential 
between the nucleons $k$ and $j$, $V_{1\pi} = v_{kj}^1 \, (\fet \tau_k \cdot \fet \tau_j )$. 
The second term in Eq.~(\ref{Tfin}) thus corresponds to the genuine contribution of the 3NF. Performing the algebra for
spin and isospin matrices, we obtain the following result:
\beqa
\label{pr1}
-i \, \delta \bar m  \, v_{ik}^2 \, [ \fet \tau_i \times \fet \tau_k]^3 \, 
 v_{kj}^1 \, (\fet \tau_k \cdot \fet \tau_j ) &=& \delta \bar m  \,  \left(
  \frac{g_A}{2 F_\pi} \right)^4 \frac{( \vec \sigma_i \cdot \vec q_{i}  ) 
(\vec \sigma_j \cdot \vec q_j  )}{(\vec q_i{} ^2 + M_{\pi}^2 )^2 ( \vec
q_j{} ^2 + M_{\pi}^2)}  \\
& & {}\times 
\bigg\{ [\vec q_i \times \vec q_j ] \cdot \vec \sigma_k  \, [ \fet \tau_i \times \fet \tau_j ]^3 
+  \vec q_i \cdot \vec q_j \, \Big[  (\fet \tau_i \cdot \fet \tau_k ) \tau_j^3  
 - (\fet \tau_i \cdot \fet \tau_j ) \tau_k^3 \Big] \nn
&& {}
- i \, \vec q_i \cdot \vec q_j  \,  [ \fet \tau_i \times \fet \tau_j ]^3  
+ i \, [\vec q_i \times \vec q_j ] \cdot \vec \sigma_k  \,  \Big[  (\fet \tau_i \cdot \fet \tau_k ) \tau_j^3  
 - (\fet \tau_i \cdot \fet \tau_j ) \tau_k^3 \Big] \bigg\}.
\nonumber
\eeqa
To get the complete expression for the 3NF we need to 
take into account the contribution of the diagram resulting from graph (a) in Fig.~\ref{fig12} by 
interchanging the ordering of the pion propagators and summing over the nucleon labels.
This leads to cancellation of the terms in the third line of Eq.~(\ref{pr1}). 
The final result agrees with $V^{11a}$ in Eq.~(\ref{3NFisosp1}).

Similarly, the scattering amplitude corresponding to the graph (b) in Fig.~\ref{fig12}
is:
\beq
\label{Tprom1}
T^{12b} = - i \, 2 \, \delta \bar m \, v_{ik}^2 \, [ \fet \tau_i \times \fet \tau_k]^3 \, \left( \frac{{\vec p_i {}'}^2}{2 m}
-  \frac{\vec p_i {}^2}{2 m} \right) \, \frac{1}{\delta  T + i \epsilon }  \, \Big[ C_S + C_T \, ( \vec \sigma_j \cdot
\vec \sigma_k ) \Big]\,.
\eeq
Notice that we use the following Feynman rule for contact terms: $(-i ) \, [ C_S + C_T  \, ( \vec \sigma_j \cdot
\vec \sigma_k )]$, which leads to the 2N potential $V = C_S + C_T  \, ( \vec \sigma_j \cdot
\vec \sigma_k )$. 
We rewrite the amplitude $T^{12b}$ in the form:
\beqa 
\label{Tfin2}
T^{12b} &=& i \, \frac{\delta \bar m}{2 m}  \, v_{ik}^2 \, [ \fet \tau_i \times \fet \tau_k]^3 \,
\Big( \vec p_i {}^2 - \vec{\tilde p_k} {}^2  -{\vec p_i {}'}^2 + {\vec p_k {}'}^2 \Big)
\, \frac{1}{\delta  T + i \epsilon }  \, \Big[ C_S + C_T \, ( \vec \sigma_j \cdot
\vec \sigma_k ) \Big] \nn
&& {}  -i \, \delta \bar m  \, v_{ik}^2 \, [ \fet \tau_i \times \fet \tau_k]^3 \, 
\Big[ C_S + C_T \, ( \vec \sigma_j \cdot
\vec \sigma_k ) \Big]\,.
\eeqa
Again, the term in the first line of the above equation gives the iterative contribution to the 
amplitude, while the term in the second line is the genuine 3NF contribution:
\beqa
-i \, \delta \bar m  \, v_{ik}^2 \, [ \fet \tau_i \times \fet \tau_k]^3 \, 
\Big[ C_S + C_T \, ( \vec \sigma_j \cdot
\vec \sigma_k ) \Big]  &=&  \delta \bar m  \,  \left(
  \frac{g_A}{2 F_\pi} \right)^2 \frac{\vec \sigma_i \cdot \vec q_{i}}{(\vec q_i{} ^2 + M_{\pi}^2 )^2} 
\, [ \fet \tau_i \times \fet \tau_k]^3  \\
&& \times \Big\{ i \, C_S \, ( \vec \sigma_k \cdot \vec q_{i})
+ i \, C_T \, ( \vec \sigma_j \cdot \vec q_{i}) - C_T \, 
[ \vec \sigma_j \times \vec \sigma_k ] \cdot \vec q_i \Big\}
\nonumber
\eeqa
The first two terms in the curly brackets cancel against the contribution of the time--reversed 
diagram, so that the final result agrees with $V^{11b}$ in Eq.~(\ref{3NFisosp1}).

Finally, it is easy to see that the amplitude corresponding to graphs (c) and (d) in Fig.~\ref{fig12}
is reproduced by the iteration of the isospin--breaking 1PE potential in Eq.~(\ref{OPEP2}) and the leading 
isospin--symmetric 2N potential with no need for an additional 3NF. We thus have verified 
the consistency of our results for isospin--violating 1PE 2N force and 3NFs.

\section{Summary and outlook}
\label{sec:sum}

The pertinent findings of this study can be summarized as follows:
\begin{itemize}
\item[1)]
We have applied the method of unitary transformation to study 
the isospin--breaking NN forces.
We have derived \emph{all} forces (in the absence of virtual photons) up to
$\nu = 5$ in the chiral expansion.  Together with the 3NFs worked out in 
Ref.~\cite{Epelbaum:2004xf}, 
this completely specifies the isospin--breaking few--nucleon  forces up to this order.
\item[2)]
The isospin--violating one--pion exchange potential is a combination of terms
of order $\nu = 2, 3,4$ and $5$. It is specified in 
Eqs.~(\ref{OPEnijm}), (\ref{OPEP2}),  (\ref{OPEP1})
and (\ref{OPEP3}). In particular, we have
reproduced the expressions for the 1PE potential, class IV, found recently in 
Ref.~\cite{Friar:2004ca}. 
%*EE
%We also find an additional interaction $\propto 
%(\delta \bar m )^2$  not considered in that reference (but e.g.~in thesis of 
%Stoks \cite{Stoks:1990bb}).
\item[3)]
The results for the isospin--violating 2PE potential in momentum space are
given in Eqs.~(\ref{VC4}), (\ref{WC4}), (\ref{VT5}) and (\ref{WC5}). 
The corresponding expressions in coordinate space are discussed in detail 
in section \ref{sec:TPEsize}. In particular, we have 
reproduced the expressions for the 2PE potential at order $\nu = 4$ worked out 
in Refs.~\cite{Niskanen:2001aj,Friar:2003yv}.
The main new result is the 2PE potential at order $\nu = 5$ which has not been 
considered in EFT before. We find that the derived subleading CSB 2PE at order $\nu = 5$ 
gives numerically the \emph{dominant} contribution to the CSB
TPE potential, cf. Figs.~\ref{fig8},\ref{fig9}. 
As in the isospin--conserving case, this effect can be traced
back to the large magnitude of the dimension two pion--nucleon LECs $c_2, c_3$
and $c_4$ (which is well understood, see \cite{Bernard:1996gq}). 
\item[4)]
The contact interactions appear at orders $\nu = 3,4,5$ and  are listed in 
Eq.~(\ref{VC}). These are parameterized in terms of the LECs $\tilde \beta_i$ and $\beta_i$. 
\item[5)]We have also demonstrated explicitely  that our results for the
  isospin--violating 2N and 3N forces are consistent with each other.
\end{itemize}

These results pave the way for new precision studies. To make this point more
transparent, we reiterate that the finite--range part of the potential, i.e. the
1PE and 2PE pieces, depend on the mass shifts 
$\delta  \bar M_\pi^2$ and $\delta \bar m$ which are well known, and the
strong nucleon mass shift  $(\delta \bar m )^{\rm str}$ which is less well
known. For the latter quantity, one can use the value based on the 
Cottingham sum rule, see Eq.~(\ref{m_shift2}). In addition, the pion--nucleon 
coupling constants $f_p^2$,  $f_n^2$ and $f_c^2$ are not well known at
present.  In principle, these constants can be extracted from 
an independent partial wave analysis of the {\it pp} and {\it np} data, 
such as the new Nijmegen PWA \cite{Rentmeester:2004pi}. Provided such an 
extraction is possible, the finite--range part of the isospin--violating 
nucleon force is completely determined.  One can then try to fix the values of
the LECs $\tilde \beta_i$ and $\beta_i$ from the 
low partial waves in the {\it np} and {\it pp} systems in order to make 
predictions for the corresponding  {\it nn} partial waves. It remains to be
seen whether this ambitious program can be carried out. Also,
one should keep in mind that the isospin--conserving NN forces have so far
only  been worked out up to order $\nu = 4$ in the power counting. 

%%%%%%%%%%%%%%%%%%%%%%%%%%%%%%%%%%%%%%%%%%%%%%%%%%%%%%%%%%%%%%%%%%%%%%%%%%%%%%%%%
\section*{Acknowledgments}

E.E.~would like to thank Prof.~W.~Gl\"ockle for helpful discussions and  warm 
hospitality during his stay in Bochum, where part of this work has been done. 
This work has been supported by the 
U.S.~Department of Energy Contract No.~DE-AC05-84ER40150 under which the 
Southeastern Universities Research Association (SURA) operates the Thomas Jefferson 
National Accelerator Facility and by the Deutsche Forschungsgemeinschaft
through funds provided to the SFB/TR 16 ``Subnuclear Structure of Matter''. 
This research is part of the EU Integrated Infrastructure Initiative 
Hadron Physics Project under contract number RII3-CT-2004-506078.

\bigskip
%%%%%%%%%%%%%%%%%%%%%%%%%%%%%%%%%%%%%%%%%%%%%%%%%%%%%%%%%%%%%%%%%%%%%%%%%%%%%%%%%%%
\bibliographystyle{h-physrev4}
\bibliography{/home/evgeny/refs}

\end{document}